\documentclass[twocolumn]{aastex62}

\usepackage{amssymb,amsmath}
\usepackage[inline]{enumitem} 

\newcommand{\alfven}{Alfv{\'e}n\,}

\newcommand{\dm}{ $$_{d}$$ }

\newcommand{\pder}[2][]{\frac{\partial#1}{\partial#2}}
\newcommand{\lder}[2][]{\frac{D #1}{D #2}}

\received{}
\revised{}
\accepted{}
\submitjournal{ApJ}

\shorttitle{}
\shortauthors{Syntelis et al.}

\begin{document}

\title{Successful and Failed Flux Tube Emergence in the Solar Interior}

\correspondingauthor{P. Syntelis}
\email{ps84@st-andrews.ac.uk}

\author{P. Syntelis}
\affiliation{St Andrews University,
Mathematics Institute,
St Andrews KY16 9SS,
UK}

\author{V. Archontis}
\affiliation{St Andrews University,
Mathematics Institute,
St Andrews KY16 9SS,
UK}

\author{A. Hood}
\affiliation{St Andrews University,
Mathematics Institute,
St Andrews KY16 9SS,
UK}

\begin{abstract}
We report on our three-dimensional (3D) magnetohydrodynamic (MHD) simulations of cylindrical weakly twisted flux tubes emerging from 18 Mm below the photosphere. We perform a parametric study, by varying the initial magnetic field strength ($B_0$), radius ($R$), twist ($\alpha)$ and length of the emerging part of the flux tube ($\lambda$) to investigate how these parameters affect the transfer of the magnetic field from the convection zone to the photosphere. We show that the efficiency of emergence at the photosphere (i.e. how strong the photospheric field will be in comparison to $B_0$) depends not only on the $B_0$ but also the morphology of the emerging field and the twist.  We show that parameters such as $B_0$ and magnetic flux cannot alone determine whether a flux tube will emerge to the solar surface.  For instance, high-$B_0$ (weak-$B_0$) fields may fail (succeed) to emerge at the photosphere, depending on their geometrical properties.  We also show that the photospheric magnetic field strength can vary greatly for flux tubes with the same $B_0$ but different geometric properties. Moreover, in some cases we have found scaling laws, whereby the magnetic field strength scales with the local density as $B\propto \rho^\kappa$, where $\kappa \approx 1$ deeper in the convection zone and $\kappa <1$, close to the photosphere. The transition between the two values occurs approximately when the local pressure scale ($H_p$) becomes comparable to the diameter of the flux tube ($H_p\approx2R$). We derive forms to explain how and when these scaling laws appear and compare them with the numerical simulations. 
\end{abstract}


\section{Introduction}

It is believed that the origin of the magnetic field of the Sun is associated with the existence of a dynamo mechanism operating around the base of the deep convection zone \citep[][]{Parker_1955b}. 
The magnetic fields rise from the 200~Mm deep convection zone towards the photosphere due to buoyancy \citep[][]{Parker_1955}, where they can emerge and form a variety of magnetic structures (from small scale pores to large scale active regions). 
The expansion of the flux tubes during their emergence within the solar interior, depends mostly on the local density of the convection zone.
However, the density inside the convection zone drops by six orders of magnitude, and the main density decrease occurs mostly in the upper convection zone. For example, the density drops by approximately $10^4$ in the upper 20~Mm of the convection zone, of which a $10^3$ drop occurs only in the upper 10~Mm. 
So, the local pressure scale height ($H_p$) is large and decreases slowly at larger depths.
Therefore, deeper in the solar interior, the motion of magnetic fields (e.g. a flux tube) is affected less by pressure variations than near the surface.
This allows the flux emergence process there to be studied using either the thin-flux tube approximation \citep[e.g.][]{Spruit_1981,Caligari_etal1995, Fan_etal1993, Weber_etal2011} or the anelastic MHD approximation \citep[e.g.][]{Brun_etal2004, Fan_2008, Jouve_etal2009, Fan_etal2014}.
Closer to the photosphere, on the other hand,  $H_p$ becomes small and decreases rapidly. Hence, the size of the emerging structures can become comparable to $H_p$ and full 3D compressible MHD is needed in order to study flux emergence in the upper convection zone.

\begin{figure*}
\centering
\includegraphics[width=\textwidth]{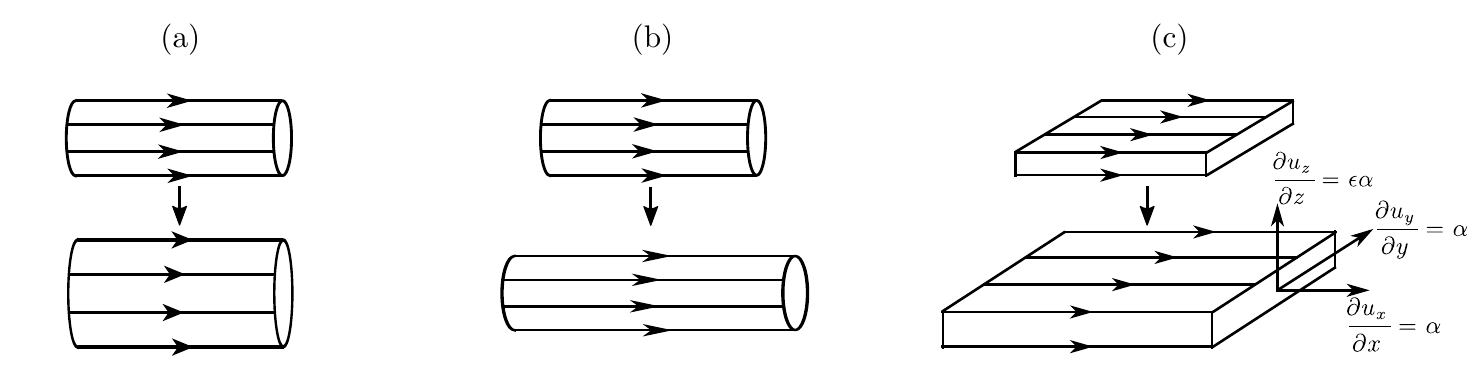}
\caption{
The three different cases of flux tube expansion discussed in the Introduction. Panel \textbf{(a)} shows the expansion along the cross section of cylindrical flux tube. Panel \textbf{(b)} shows the expansion along the length of the flux tube. Panels \textbf{(c)} shows the expansion of a horizontal magnetic field due to the presence of velocity gradients.
}
\label{fig:introduction}
\end{figure*}

\citet{Toriumi_etal2010} performed 2D MHD simulations of a magnetic flux sheet positioned at $z=-20$~Mm below the photosphere. They reported that in order for the magnetic field to emerge into the photosphere and above, its flux needs to be $10^{21}-10^{22}$~Mx. However, these fluxes were derived by assuming the length of the magnetic flux sheet along the third dimension.
\citet{Toriumi_etal2013} performed 3D MHD simulations of a magnetic flux tube originating from the same depth. They varied the initial magnetic field strength, twist and length of the buoyant part of the flux tube. 
They found that for higher values of the magnetic field strength and twist, the flux tube emerges faster inside the solar interior, and expands more dynamically above the photosphere. 
On the other hand, the flux tube emerges faster inside the solar interior, but expands less dynamically above the photosphere, when the buoyant part of the flux tube is longer. 
The above results are important to understand the emergence process of flux tubes in the solar interior. However, many questions remain open. For example, how the parameters of the initial sub-photospheric field affect the amount of flux emerging at the photosphere is still unknown.

Another interesting question is how the magnetic field strength ($B$) scales with the local plasma density ($\rho$) during the emergence of the flux tube within the convection zone.
A simple scaling law can be derived if we assume a flux tube with a uniform axial magnetic field of strength $B$ and then vary its cross section ($A$), while keeping its length ($L$) constant (Fig.~\ref{fig:introduction}a). From conservation of mass ($M=AL\rho$) and magnetic flux ($\Phi=AB$), it is easy to show that $B\propto\rho$ (or $B\propto\rho^{\kappa}$ with $\kappa=1$).

Another scaling law can be derived if we vary the length of the flux tube while keeping its cross section constant  (Fig.~\ref{fig:introduction}b). Conservation of mass and flux suggests that $B$ and $\rho$ will depend on the length of the flux tube. Useful information about the scaling can be derived by assuming that $B\propto\rho^{\kappa}$. Then, $\kappa$ becomes constrained to $0<\kappa<1$ \citep{Pinto_etal2013}. 

Finally, the scaling of the magnetic field strength with the local
plasma density can be affected by the action of velocity gradients on the 
magnetic field. To show that, \citet{Cheung_etal2010} assumed a horizontal 
magnetic field, $\mathbf{B} = B \; \mathbf{\hat{x}}$, (Fig.~\ref{fig:introduction}c). 
This field was then distorted by an asymmetric velocity gradient, defined by:
\begin{align}
    \pder[v_x]{x} =  \alpha, \quad
    \pder[v_y]{y} = \alpha, \quad
    \pder[v_z]{z} = \epsilon \alpha,
    \label{eq:velocity_cheung}
\end{align}
where $\alpha$ is the horizontal expansion rate and $\epsilon$ is a parameter describing the asymmetry of the flow in the vertical direction. Combining the ideal induction and the continuity equations,
\begin{align}
    \lder[\mathbf{B}]{t} &= 
    - ( \mathbf{\nabla} \cdot \mathbf{v} ) \textbf{B}
    + ( \textbf{B} \cdot \mathbf{\nabla} ) \textbf{v} \label{eq:induction} \\
    \lder[\rho]{t} &= 
    - \rho ( \mathbf{\nabla} \cdot \mathbf{v} ), \label{eq:continuity}
\end{align}
they found that the scaling of $B$ with $\rho$ is indeed affected by the velocity gradients and that the power $\kappa$ depends on the degree of the asymmetry of the velocity gradients as 
\begin{align}
    \kappa=\frac{1+\epsilon}{2+\epsilon}. 
    \label{eq:kappa_cheung}
\end{align} 
For a purely horizontal expansion ($\epsilon=0$), $\kappa=0.5$ and for expansion transverse to the field ($\epsilon>>1$), $\kappa=1$, as expected from the conservation of flux and mass (as in Fig.~\ref{fig:introduction}a). For isotropic expansion ($\epsilon=1$), they found $\kappa=2/3$.

But which scaling is more suitable at different depths inside the convection zone?
In the deeper parts of the convection zone, the local pressure scale height is large. The characteristic length of the flux tube's cross section (e.g. radius, $r$) is $r<<H_p$ and the characteristic length of the emerging part (e.g. an axial perturbation, $l$) is $l>>H_p$ (i.e. a thin flux tube). 
So, the flux tube's axial expansion would be small and its cross sectional expansion would be gradual and symmetric. As a result, velocity gradients along the axis would be small and the scaling of the magnetic field with the local density should follow $\kappa=1$.

\citet{Pinto_etal2013} studied the emergence of twisted flux tubes in a global dynamo model, using 3D anelastic MHD simulations. 
They found that $B\propto\rho^\kappa$, $\kappa=0.998\pm0.001$.
Similar behaviour was found in cases without a dynamo. 
When the emerging field was less buoyant, they found steeper, but similar slopes. Overall, they found values of $0.998<\kappa<1.002$ during the emergence of the flux tubes from 0.8~$R_{\odot}$ to 0.92~$R_{\odot}$. These results suggested that the poloidal component of the magnetic field dominated over the toroidal component and that the perturbations along the axis of the tube where indeed small.

In the upper parts of the convection zone, the length of the flux tube can increase significantly as the $\Omega$-loop shaped flux tube rises towards the photosphere.
Moreover, the flux tube expands radially, as its cross section becomes comparable to $H_p$.
Close to the photosphere, the flux tube experiences a significant horizontal expansion \citep[][]{Spruit_etal1987} due to the rapid decrease of $H_p$. The flux tube cannot emerge above the photosphere until its magnetic forces dominate over the gas pressure forces and trigger a magnetic buoyancy instability \citep{Acheson1979,Archontis_etal2004}.
So, when the flux tube is underneath the photosphere, it becomes compressed and further expands horizontally, increasing its magnetic field strength until it is large enough to trigger the buoyancy instability.
In addition to the above, the velocity gradients of the convective flows are larger closer to the photosphere than deeper in the solar interior.
All the above should lead to a decrease of $\kappa$. Thus, the scaling of the magnetic field with the local density should follow $\kappa<1$ in the upper convection zone.

\citet{Cheung_etal2010} compared their analytical result on $\kappa$, with a 3D radiative MHD simulation of the emergence of a toroidal flux tube, positioned $7.5$~Mm below the photosphere, inside a convective layer. 
They found that a value of $\kappa=0.5$ for the scaling of the magnetic field strength with the local density.

\citet{Cheung_etal2014} suggested that the transition from $\kappa\approx1$ \citep[similar to][]{Pinto_etal2013} to $\kappa<1$ occurs somewhere during the rise of a flux tube from the deeper parts of the convection zone to the surface. 

In this paper, we address a series of questions on the emergence of flux tubes from the convection zone to the photosphere. 
For this, we use 3D resistive and compressible MHD and assume a horizontal flux tube positioned at 18~Mm below the photosphere. 
The free parameters of our model are a) the initial magnetic field strength at the center of the flux tube, b) the twist, c) the radius and d) the length of the buoyant part of the flux tube.
We perform a detailed parametric study to identify i) how $\kappa$ behaves when each of these parameters are varied, ii) where does the transition from $\kappa\approx1$ to $\kappa<1$ occur, 
iii) what is the efficiency of emergence, namely what is the ratio of the photospheric field strength to the initial field strength, and iv) how to use the above in order to understand the initial conditions leading to ``successful'' flux emergence above the photosphere.
Furthermore, we derive analytically the conditions under which $\kappa$ is constant.

In Sec.~\ref{sec:initial_conditions}, we present the model and the initial conditions. 
In Sec.~\ref{sec:fluxes}, we vary only the magnetic field strength and radius of the flux tube, in order to explore the parameter space and identify combinations of those two parameters leading to ``successful'' emergence of magnetic field above the photosphere.
In Sec.~\ref{sec:scalings}, we describe analytically conditions  under which $B$ scales with $\rho$, and compare our analysis with a numerical simulation of a ``successful'' emergence.
In Sec.~\ref{sec:parametric}, we focus on one of the ``successfull'' emergence cases of Sec.~\ref{sec:fluxes} and perform a large parametric study to identify how each parameter affects the emergence of the field (magnetic field strength (Sec.~\ref{sec:field_strength}), radius (Sec.~\ref{sec:radius}), length of the buoyant part (Sec.~\ref{sec:lambda}) and twist (Sec.~\ref{sec:alpha}) ).
In Sec.~\ref{sec:all_cases} we discuss all the results together, and present a ``border-line'' case that separates the ``successful'' and the ``failed'' emergence cases.
In Sec.~\ref{sec:conclusions} we summarize and discuss our results.

\section{Numerical Setup}
\label{sec:initial_conditions}

\begin{figure}
\centering
\includegraphics[width=\columnwidth]{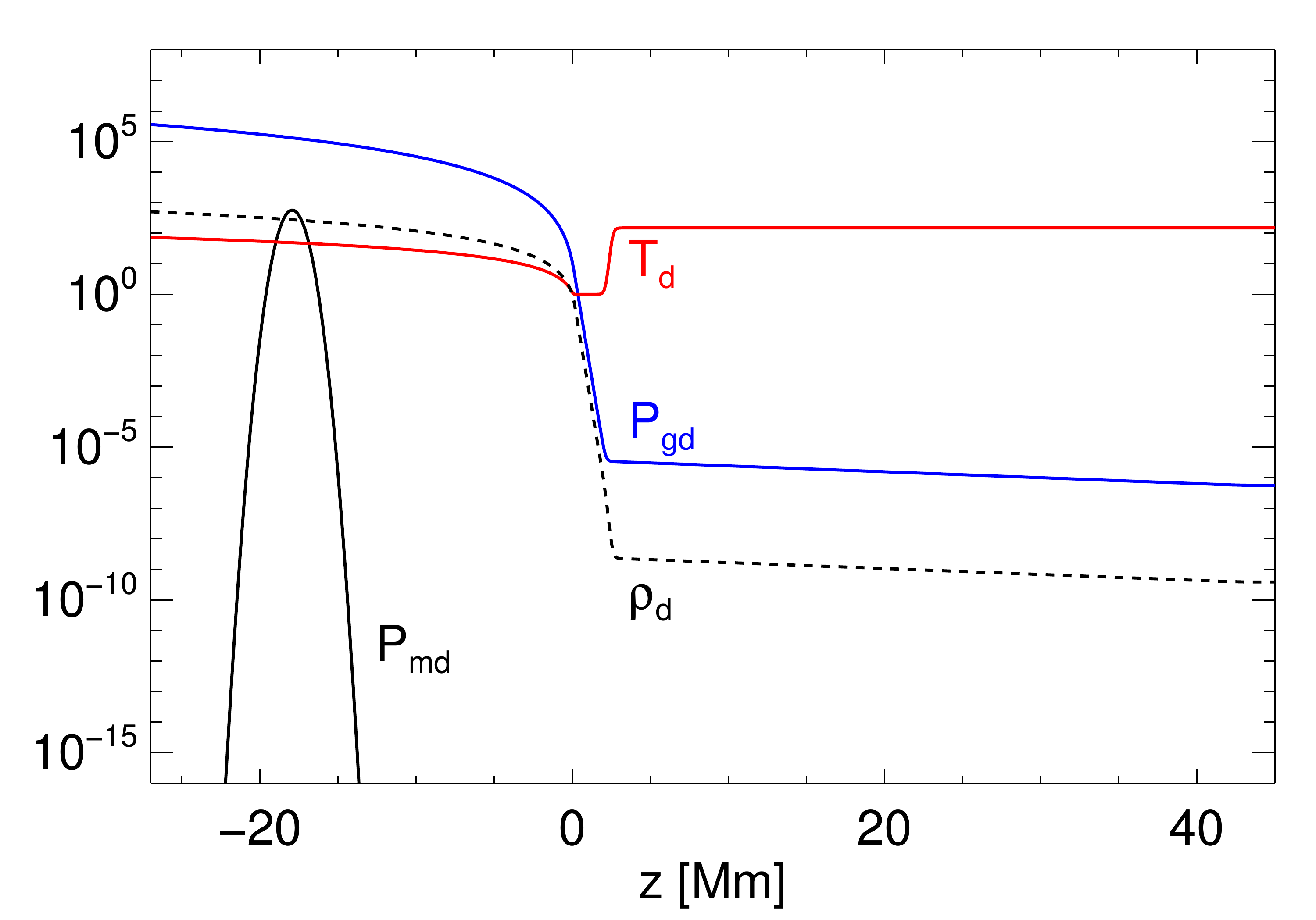}
\caption{
Initial stratification of the atmosphere in dimensionless units (temperature ($T$), density ($\rho$), magnetic pressure ($P_m$ of the case~10, Table~\ref{tab:fluxes_b} flux rope) and gas pressure ($P_g$)).
}
\label{fig:stratification}
\end{figure}

\begin{deluxetable}{cccccc}
\tablenum{1}
\tablecaption{ }
\label{tab:fluxes_b}
\tablewidth{0pt}
\tablehead{
\colhead{Case} & \colhead{B$_0\dm$} & \colhead{$R\dm$ } & \colhead{$\lambda\dm$} & \colhead{$\alpha\dm$}  &  \colhead{$\Phi$} \\
\colhead{} & \colhead{($\times B_c$)} & \colhead{($\times H_c$)} & \colhead{($\times H_c$)} & \colhead{($\times H_c^{-1}$)}  &  \colhead{(Mx)}
}
\startdata
	1*    & 3.4 & 3.2   & 100           & 0.1           & $1.1\times10^{19}$        \\ 
	2*    & 3.4 & 5     & 100           & 0.1           & $2.6\times10^{19}$        \\ 
	3*    & 3.4 & 7.6   & 100           & 0.1           & $6.0\times10^{19}$        \\
	4*    & 3.4 & 10.1  & 100           & 0.1           & $1.1\times10^{20}$        \\ 
	5*    & 17  & 3.2   & 100           & 0.1           & $5.3\times10^{19}$        \\ 
	6*    & 17  & 5     & 100           & 0.1           & $1.3\times10^{20}$        \\ 
	7*    & 17  & 7.6   & 100           & 0.1           & $3\times10^{20}$          \\ 
	8     & 17  & 10.1  & 100           & 0.1           & $5.3\times10^{20}$        \\
	9*    & 34  & 3.2   & 100           & 0.1           & $1.1\times10^{20}$        \\ 
	10    & 34  & 5     & 100           & 0.1           & $2.6\times10^{20}$        \\ 
	11    & 34  & 7.6   & 100           & 0.1           & $6.0\times10^{20}$        \\ 
	12    & 34  & 10.1  & 100           & 0.1           & $1.1\times10^{21}$        \\ 
	13    & 340 & 3.2   & 100           & 0.1           & $1.1\times10^{21}$        \\ 
	14    & 340 & 5     & 100           & 0.1           & $2.6\times10^{21}$        \\ 
	15    & 340 & 7.6   & 100           & 0.1           & $6.0\times10^{21}$        \\ 
	16    & 340 & 10.1  & 100           & 0.1           & $1.1\times10^{22}$        \\
\enddata
\tablecomments{
The values of the initial parameters of the simulations used to produce Fig.~\ref{fig:flux_B}. The cases denoted with an asterisk represent ``failed'' emergence above the photosphere. The other cases represent ``successful'' emergence above the photosphere.
}
\end{deluxetable}

\begin{deluxetable}{cccccc}
\tablenum{2}
\tablecaption{ }
\label{tab:scaling}
\tablewidth{0pt}
\tablehead{
\colhead{Case} & \colhead{B$_0\dm$} & \colhead{$R\dm$ } & \colhead{$\lambda\dm$} & \colhead{$\alpha\dm$}  &  \colhead{$\Phi$} \\
\colhead{} & \colhead{($\times B_c$)} & \colhead{($\times H_c$)} & \colhead{($\times H_c$)} & \colhead{($\times H_c^{-1}$)}  &  \colhead{(Mx)}
}
\startdata
	1*    & 17      & 5     & 100           & 0.1      & $1.3\times10^{20}$      \\ 
	2*    & 24      & 5     & 100           & 0.1      & $1.8\times10^{20}$      \\ 
	3     & 34      & 5     & 100           & 0.1      & $2.6\times10^{20}$      \\
	4     & 68      & 5     & 100           & 0.1      & $5.1\times10^{20}$      \\ 
	5     & 34      & 3.2   & 100           & 0.1      & $1.1\times10^{20}$      \\ 
	6     & 34      & 7.6   & 100           & 0.1      & $6.0\times10^{20}$      \\ 
	7     & 34      & 10.1  & 100           & 0.1      & $1.1\times10^{21}$      \\ 
	8*    & 34      & 5     & 20            & 0.1      & $2.6\times10^{20}$      \\
	9     & 34      & 5     & 35            & 0.1      & $2.6\times10^{20}$      \\ 
	10    & 34      & 5     & 50            & 0.1      & $2.6\times10^{20}$      \\ 
	11    & 34      & 5     & 100           & 0.15     & $2.6\times10^{20}$      \\ 
	12    & 34      & 5     & 100           & 0.25     & $2.6\times10^{20}$      \\ 
	13    & 34      & 5     & 20            & 0.11     & $2.6\times10^{20}$      \\ 
	14    & 34      & 5     & 20            & 0.15     & $2.6\times10^{20}$      \\ 
	15    & 34      & 5     & 20            & 0.25     & $2.6\times10^{20}$      \\ \hline
	16*   & 34      & 3.2   & 50            & 0.1      & $1.1\times10^{20}$      \\
	17*   & 44      & 5     & 10            & 0.1      & $3.4\times10^{20}$      \\ 
	18*   & 24      & 5     & 20            & 0.1      & $1.8\times10^{20}$      \\ 
	19*   & 24      & 5     & 20            & 0.25     & $1.8\times10^{20}$      \\ 
	20    & 24      & 7.6   & 100           & 0.1      & $4.3\times10^{20}$      \\
	21*   & 17      & 7.6   & 100           & 0.1      & $3.0\times10^{20}$      \\ 
\enddata
\tablecomments{
Cases 1-15 show the initial parameters of the simulations used to produce Fig.~\ref{fig:h_t_profiles},~\ref{fig:scaling}.
Cases 1-21 show the initial parameters of the simulations used to produce Fig.~\ref{fig:scaling_all}.
The cases denoted with an asterisk represent ``failed'' emergence above the photosphere. The other cases represent ``successful'' emergence above the photosphere.
}
\end{deluxetable}
To perform the simulations, we numerically solve the 3D time-dependent, resistive, compressible MHD equations in Cartesian geometry using the Lare3D code of \citet{Arber_etal2001}. The equations in dimensionless form are:
\begin{align}
 \pder[\rho]{t}  =&  -\nabla \cdot (\rho \mathbf{v}) , \\
 \pder[(\rho \mathbf{v})]{t} =& -  \nabla \cdot (\rho \mathbf{v \otimes v}) 
                                 + (\nabla \times \mathbf{B}) \times \mathbf{B} - \nabla P \\
                               & - \rho g_0 \mathbf{\hat{z}} + \mathbf{S}_{visc}  , \\
\pder[\rho \epsilon]{t} = & - \nabla \cdot (\rho \epsilon \mathbf{v}) 
                            -P \nabla \cdot \mathbf{v}+ Q_{joule} 
                            + Q_{visc}, \\
\pder[\mathbf{B}]{t} =& \nabla \times (\mathbf{v}\times \mathbf{B})+ \eta \nabla^2 \mathbf{B},\\
\epsilon  =&\frac{P}{(\gamma -1)\rho},
\end{align}
where $\rho$, $\mathbf{v}$, $\mathbf{B}$ and P are density, velocity vector, magnetic field vector and gas pressure. Gravity is $g_0=274$~m s$^{-2}$. We assume a perfect gas with ratios of specific heat $\gamma=5/3$. Viscosity is included through:
\begin{align}
\mathbf{S}_{visc}= \pder[ ]{x_j}\left(\sigma_{ij} + \sigma_{ij}^{shock}\right) \mathbf{\hat{e}}_i,
\end{align}
where $\sigma_{ij} = 2 \nu \left( \varepsilon_{ij} - \frac{1}{3}\delta_{ij} \nabla \cdot \mathbf{v} \right)$ is the viscous stress tensor and 
$\sigma_{ij}^{shock} = \rho l (\nu_1 c_{ms}  + \nu_2  l |s| ) \left( \varepsilon_{ij} - \frac{1}{3}\delta_{ij} \nabla \cdot \mathbf{v} \right)$
 is the shock tensor. 
In these tensors, 
$\varepsilon_{ij} = \frac{1}{2} \left( \pder[v_i]{x_j} + \pder[v_j]{x_i} \right)$
is the strain rate tensor, $\delta_{ij}$ is the Kronecker delta, $l$ is the distance across a grid cell in the direction normal to the shock front, $|s|$ is the rate of the strain tensor in the direction normal to the shock front and $c_{ms}=\sqrt{c_s^2+v_A^2}$ is the magnetosonic speed, with $c_s$ being the sound speed and $v_A$ being the \alfven speed \citep[more details in e.g.][]{Arber_etal2001,Bareford_Hood2015}.
The viscosity coefficients take the values $\nu=622$~kg m$^{-1}$ s$^{-1}$ ($0.01$ in non-dimensional units), and $\nu_1=0.1$ and $\nu_2=0.5$ (in non-dimensional units).
Viscous heating is added through $Q_{visc} = \varepsilon_{ij}(\sigma_{ij}+\sigma_{ij}^{shock})$.

We use constant explicit resistivity of $\eta=0.01$ (non-dimensional units). Joule dissipation is added through $Q_\mathrm{joule}=\eta j^2$.
The normalization is based on the photospheric values of density $\rho_\mathrm{c}=1.67 \times 10^{-7}\ \mathrm{g}\ \mathrm{cm}^{-3}$, length $H_\mathrm{c}=180 \ \mathrm{km}$ and magnetic field strength $B_\mathrm{c}=300 \ \mathrm{G}$. 
From these we get temperature $T_\mathrm{c}=6230~\mathrm{K}$, pressure $P_\mathrm{c}=7.16\times 10^3\ \mathrm{erg}\ \mathrm{cm}^{-3}$, velocity $v_\mathrm{0}=2.1\ \mathrm{km} \ \mathrm{s}^{-1}$ and time $t_\mathrm{0}=85.7\ \mathrm{s}$.

The computational domain has a physical size of $72^3 \mathrm{Mm}$ on a $600^3$ uniform grid. We assume periodic boundary conditions in the $y$ direction. Open boundary conditions are used in the $x$ direction. Open (closed) boundary conditions are assumed and at the top (bottom) of the numerical domain.

The temperature of the atmosphere ($z>0$) follows a tangential profile, 
\begin{align}
    T(z) = T_{ph} + \frac{T_{cor}- T_{ph}}{2} 
            \left( \tanh{\frac{z-z_{cor}}{w_{tr}} +1} \right),
\end{align}
where $T_{ph}=6100$~K, $T_{cor} = 0.92$~MK, $z_{cor}=2.38$~Mm and $w_{tr}=0.18$~Mm.  
This results in an isothermal photospheric-chromospheric layer at $0 \ \mathrm{Mm} \le z < 1.8 \ \mathrm{Mm} $, a transition region  at $1.8 \ \mathrm{Mm} \le z < 3.2 \ \mathrm{Mm}$ and an isothermal coronal at $3.2 \ \mathrm{Mm} \le z < 45\ \mathrm{Mm}$.
The atmospheric density is derived by numerically solving the hydrostatic equation $dP/dz = - g\rho$, having as boundary condition $\rho_{ph}= 1.67\times10^{-7}$~g cm$^{-3}$. The atmosphere is field-free.

The solar interior ($-27\ \mathrm{Mm}\le z < 0 \ \mathrm{Mm}$) is convectively stable and in hydrostatic equilibrium.
The temperature profile of the interior increases linearly, with depth with
the constant temperature gradient given by:
\begin{align}
   T(z) = T_{ph} - \frac{\mu_m g}{k_B} \frac{\gamma-1}{\gamma}z
\end{align}
where $\mu_m= m_f m_p$ is the reduced mass, $m_p$ is proton mass,  $m_f=1.2$, and $k_B$ is Boltzmann's constant.
The density in the interior is calculated by solving the hydrostatic equation with boundary condition $\rho_{ph}$.
This stratification (sometimes with different $m_f$) is commonly used in flux emergence simulations of a fully ionized convectively stable solar interior \citep[e.g.][]{Fan_2001, Manchester_etal2004, Archontis_etal2004, Moreno-Insertis_etal2008, Toriumi_etal2011, Leake_etal2013,Syntelis_etal2015, Syntelis_etal2017}.
The initial distribution of temperature ($T$), density ($\rho$), gas pressure ($P_\mathrm{g}$) of the interior and the atmosphere is shown in Fig. \ref{fig:stratification}. The gas pressure of the interior at $-20$~Mm ($-10$~Mm) is 1.1$\times10^4$ (3$\times10^3$) larger than the photospheric one.

We place a cylindrical flux tube at $z=-18$~Mm, oriented along the $y$-axis. 
The magnetic field of the flux tube is defined as:
\begin{align}
B_{y} &=B_\mathrm{0} \exp(-r^2/R^2), \label{eq:by}\\
B_{\phi} &= \alpha r B_{y}  \label{eq:bphi}
\end{align}
where $R$ is a measure of the flux tube's radius, $r$ the radial distance from the flux tube's axis and $\alpha/2\pi$ is the twist per unit of length. 
The magnetic pressure ($P_m$) of a flux tube with $B_0=34$, $R=5$ is over-plotted in Fig.~\ref{fig:stratification} (black line).

The background solar interior has a pressure, temperature and density profile of $P_0$, $T_0$ and $\rho_0$. When adding the flux tube, we introduce a pressure excess due to the magnetic field. By requiring the flux tube to be in radial force balance (i.e. ($\nabla P)\cdot \mathbf{\hat{e}_r} = (\mathbf{j}\times\mathbf{B} )\cdot \mathbf{\hat{e}_r}$), we find the excess pressure $P_{exc}$ to be \citep[see details in][]{Murray_etal2006}:
\begin{align}
    P_{exc} = \frac{1}{2\mu}\left[ \alpha^2 \left( \frac{R^2}{2} -r^2 \right) -1 \right] B_y^2.
\end{align}
So, to set the flux tube in force balance with the background, we set the gas pressure inside the flux tube ($P_i$) to be $P_i=P_0 - P_{exc}$.
To initiate the flux tube emergence, we assume the flux tube is in thermal equilibrium with the background ($T_i=T_0$) and this leads to a difference of density of $\Delta\rho = \rho_i-\rho_0=- \rho_0 P_{exc}/P_0$ between the flux tube interior and the non-magnetized background plasma (density deficit), which makes the flux tube buoyant. To avoid emerging the whole length of the flux tube, we reduce the density deficit towards the flanks of the flux tube by \citep{Fan_2001}:
\begin{align}
    \Delta\rho = - \rho_0 \frac{P_{exc}}{P_0} e^{-y^2/\lambda^2},
    \label{eq:density_deficit}
\end{align}
where $\lambda$ is thus a measure of the length of the buoyant part of the flux tube. The above ensures that the middle part of the flux tube will be buoyant, while the flanks will not. Therefore, during the emergence, the flux tube will adopt an $\Omega$-loop shape during its emergence.

The values of the parameters used for our parametric numerical study are show in Tables~\ref{tab:fluxes_b},~\ref{tab:scaling}. 
From now on, we will refer to the dimensionless values of a variable using the subscript ``d'' (e.g. $B_0\dm$ will be the dimensionless initial magnetic field strength).

\section{Results}

\subsection{Magnetic flux and emergence}
\label{sec:fluxes}

\begin{figure}
\centering
\includegraphics[width=\columnwidth]{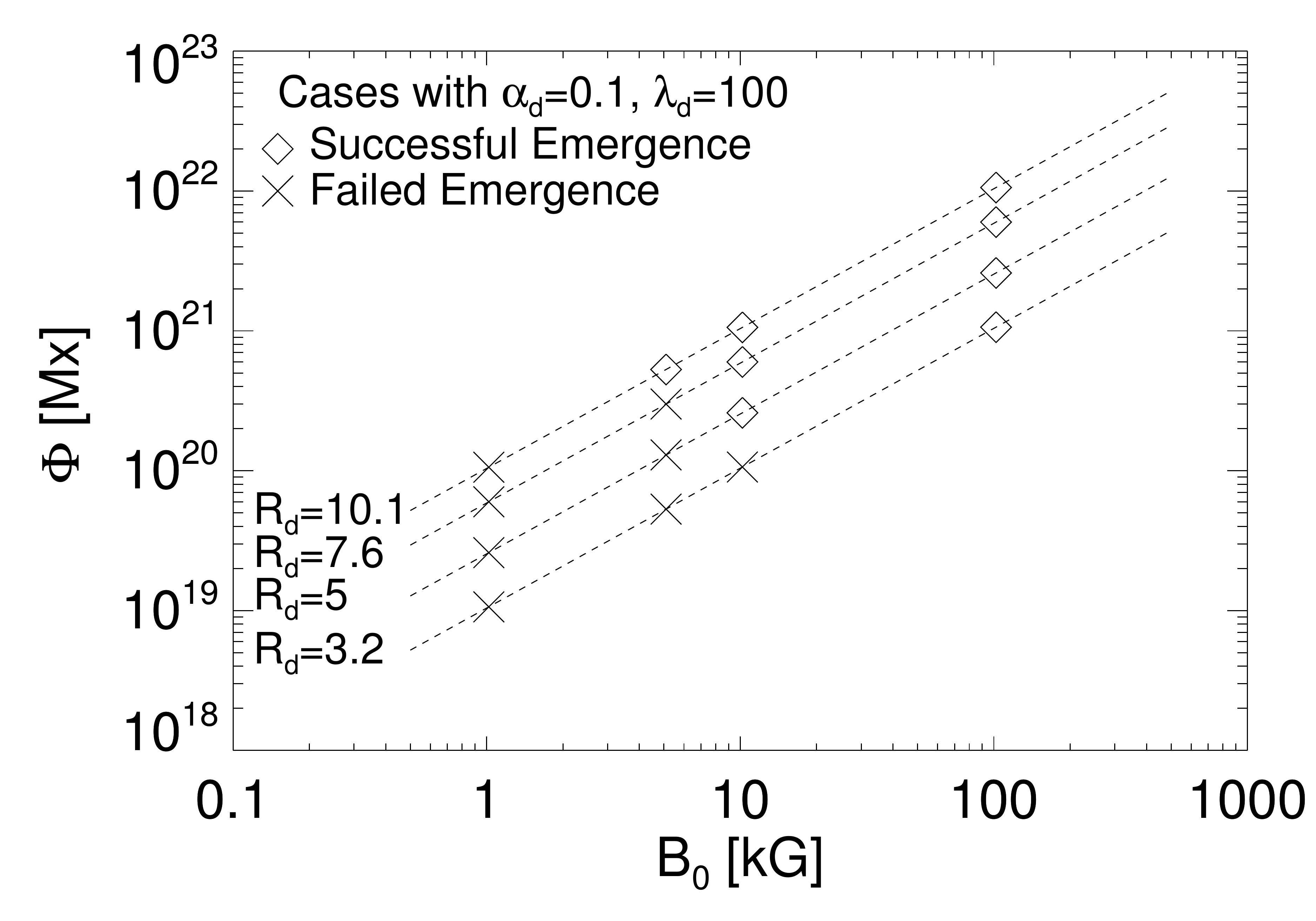}
\caption{
Plot showing the initial fluxes as a function of $B_0$ for the simulations of Table~\ref{tab:fluxes_b}. Diamonds correspond to ``successful'' emergence and ``x'' to ``failed'' emergence.  The dashed lines show flux tubes of the same radius. $R_d$ denotes the value of the radius.
}
\label{fig:flux_B}
\end{figure}

\begin{figure*}
\centering
\includegraphics[width=\textwidth]{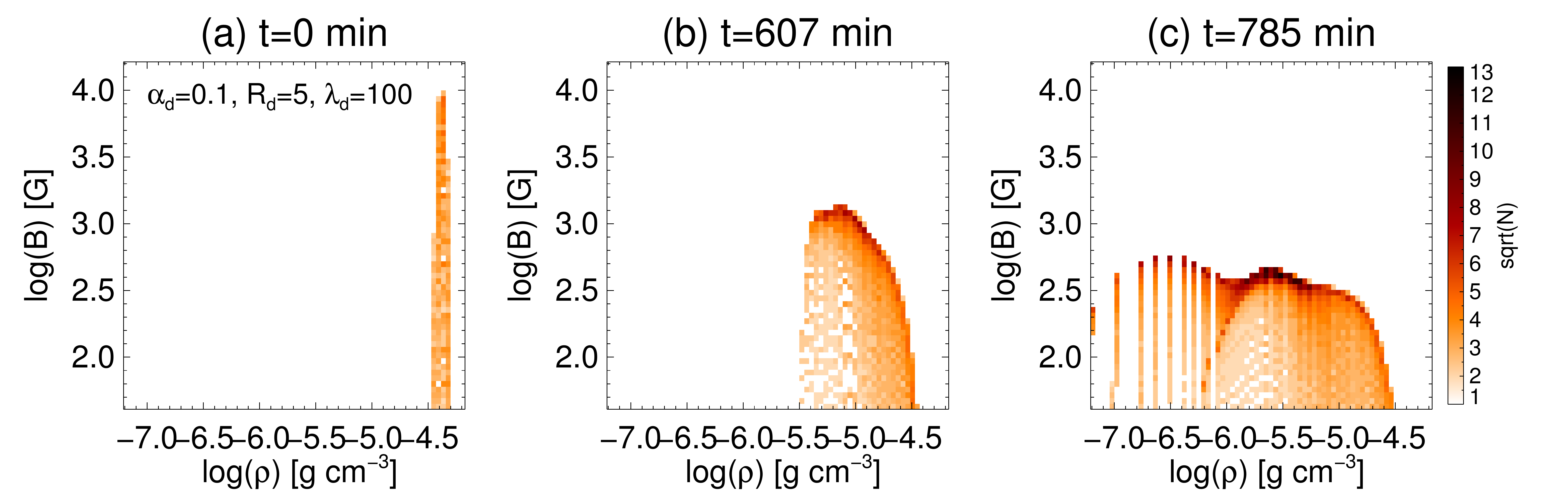}
\caption{ 
Histogram of $B$ over $\rho$ for the $R\dm=5$, $\lambda\dm=100$, $\alpha\dm=0.1$ and $B_0\dm=34$ (case 5, Table~\ref{tab:fluxes_b}). 
The values were sampled at the $xz$-midplane (plane crossing the flux tube's cross section) at (a) $t=0$~min, (b) $t=607$~min, (c) $t=785$~min.
}
\label{fig:demonsrate_hist}
\end{figure*}
First we study the emergence of flux tubes by varying their initial magnetic flux from 10$^{19}$~Mx up to 10$^{21}$~Mx. 
To change the initial magnetic flux we vary both the magnetic field strength and the radius of the flux tube. We select $B_0=1,\, 5, \, 10,\, 100$~kG ($B_0\dm=3.4,\, 17, \, 34,\, 340$) and $R=0.6,\, 0.9,\, 1.4,\, 1.8$~Mm ($R\dm=3.2,\, 5, \, 7.6,\, 10.1)$. The combination of these values produce 16 cases, shown in Table~\ref{tab:fluxes_b}. 
Fig.~\ref{fig:flux_B} shows the resulting values of the initial flux as a function of $B_0$. 
In all cases, the initial twist is low ($\alpha\dm$=0.1). In general, increasing the twist keeps the flux tube more coherent and assists the emergence process \citep[e.g.][]{Moreno-Insertis_etal1996, Murray_etal2006,Toriumi_etal2011}. We choose the length scale of the buoyant part of the flux tube to be relatively large ($\lambda\dm=100$). As a result, the apex of the flux tube will adopt a horizontal-like shape during its emergence.

\citet{Archontis_etal2004} found that a flux tube will emerge from the solar interior into the solar atmosphere when a magnetic buoyancy instability \citep{Acheson1979} is triggered. 
Before the instability is triggered, the emerging field cannot penetrate the solar surface and, instead compresses significantly just below the photosphere.
The increase in magnetic field strength at this location reduces the plasma 
$\beta$ inside the flux tube and the instability is triggered when the plasma $\beta$ drops below unity.
In our simulations, to classify a case as ``successful'' or ``failed'' emergence at and above the photosphere, we use the following criteria. 
If the buoyancy instability criterion \citep[see e.g.][]{Acheson1979, Archontis_etal2004} measured at the photosphere is satisfied and the photospheric magnetic field is at least 100~G, then we consider the emergence as ``successful''.
If the rising flux tube reaches the photosphere, but its  plasma $\beta$ remains very high (e.g. $\ge100$) and does not decrease considerably over several (at least 100) Alfv\'{e}n times, we consider the emergence as ``failed''.

Not surprisingly,  all the cases with $B_0=100$~kG emerge ``successfully'' (Fig.~\ref{fig:flux_B}). 
Most of the flux tubes with $B_0=10$~kG, manage to emerge ``successfully'', with the exception exception being the thinnest of these flux tubes ($R\dm=3.2$). The magnetic field strength at its center drops significantly, resulting in an internal $\beta\approx4000$ plasma just below the photosphere. 
For two flux tubes with same $B_0$ but different radius, at $r=R_{large}$ and $r=R_{small}$, the pressure difference between the interior and the exterior of the tube will be the same. Thus, the expansion rate of the flux tubes will be, at least initially, the same.
However, as the flux tubes expand, the cross-sectional area of the smaller flux tube grows more (as a percentage of the cross-sectional area at $t=0$).
Due to conservation of flux, the magnetic field strength of a smaller radius flux tube  will decrease more in comparison to a larger radius flux tube.
Therefore, its magnetic pressure will decrease faster and it will bring higher plasma $\beta$ material close to the photosphere. 

For $B_0=5$~kG, only the largest flux tube radius manages to emerge above the photosphere. 
All the $B_0=1$~kG cases fail to emerge.
They rise very slowly and end up reaching force balance inside the solar interior, with a very large $\beta$. In these cases, the magnetic field brought below the photosphere is very low and the buoyancy instability is never triggered.

Notice that some cases ``successfully'' emerge (e.g. $B_0=10$~kG and $R\dm=5$) while other cases with a similar flux but different $B_0$ and $R$ fail to emerge ($B_0=1$~kG and $R\dm=10.1$, $B_0=5$~kG and $R\dm=7.6$). Despite the substantial initial flux (greater than $10^{20}$~Mx), these two flux tubes are not buoyant enough to emerge ``successfully''. Therefore, we conclude that the initial magnetic flux within the rising magnetic structure cannot indicate directly whether the magnetic structure will emerge.

From Fig.~\ref{fig:flux_B} we find that in some cases an increase of $R$ (for constant $B_0$) leads to ``successful'' emergence. Such cases are the $B_0=5$~kG and $R\dm=10.1$  $B_0=10$~kG and $R\dm=5$. In Sec.~\ref{sec:parametric} we will present the results of a parametric study on $B$, $R$, $\alpha$, and $\lambda$ in the latter case, to identify how these parameters affect the emergence.
However, it is important first to show how the magnetic field strength varies with the local density, during the emergence of the flux tubes in the solar interior. This is discussed in the next section.

\subsection{Scaling of magnetic field strength with the local density}
\label{sec:scalings}

\begin{figure*}
\centering
\includegraphics[width=\textwidth]{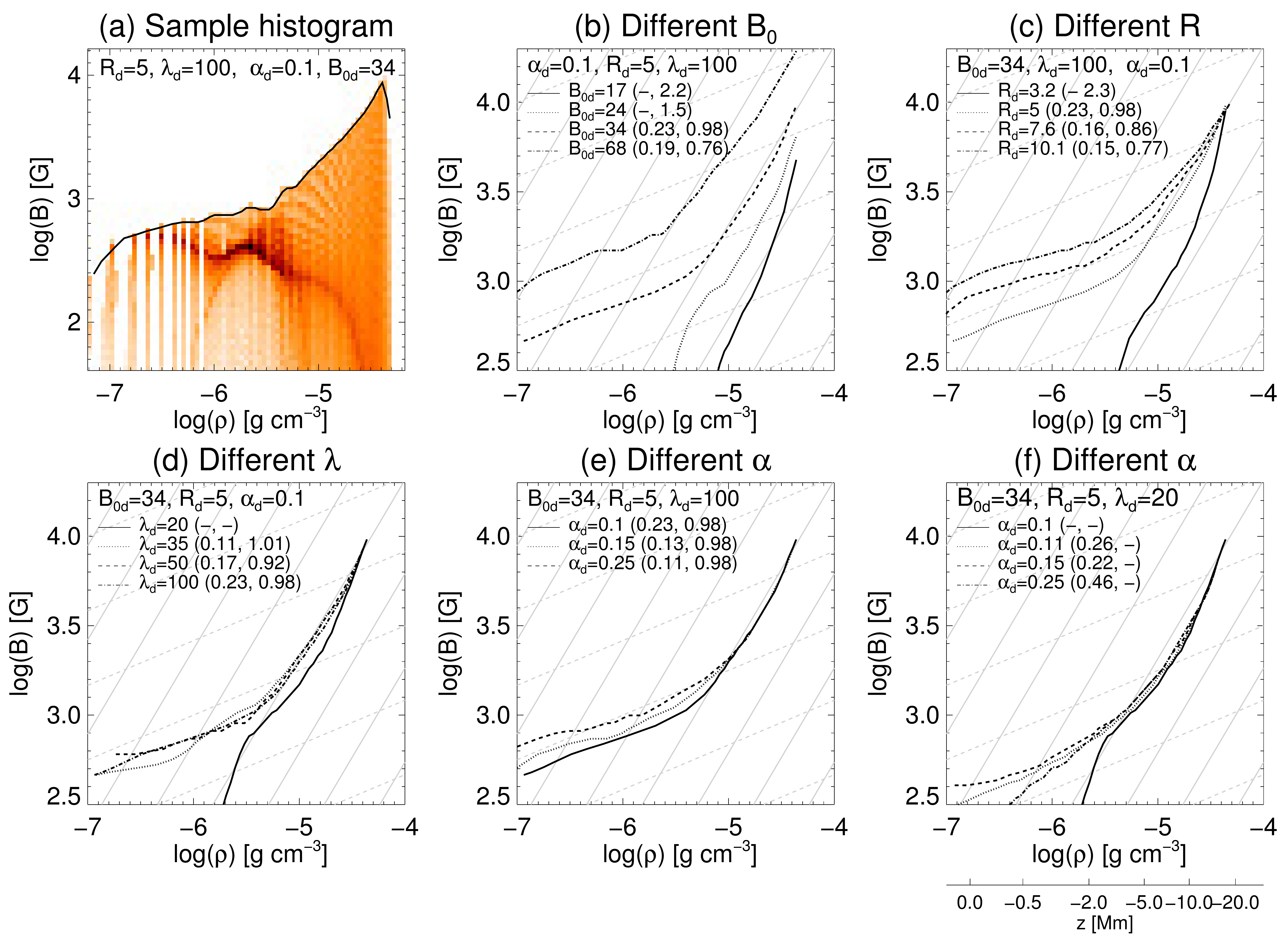}
\caption{ 
(a) Histogram of $B$ over $\rho$ for the $R\dm=5$, $\lambda\dm=100$, $\alpha\dm=0.1$ and $B_0\dm=34$ (case 5, Table~\ref{tab:fluxes_b}). 
The values were sampled at the $xz$-midplane (plane crossing the flux tube's cross section) during the whole simulation run ($t=0-950$~minutes). 
The solid black line outlining the uppermost part of the histogram is the scaling curve.
Panels (b)-(f) shows the smoothed scaling curves of cases 1-15 of Table~\ref{tab:scaling}. 
Panel (b) shows cases with different $B_0$, (c) cases with different $R$, (d) cases with different $\lambda$, (e) cases with different $\alpha$ and $\lambda\dm=100$ and (e) cases with different $\alpha$ and $\lambda\dm=20$. 
The axis below panel (f) shows the depth inside the solar interior that is equivalent to the density $x$-axis of panels (b)-(f).
The legends in each panel show the specific parameters of each simulation. 
The value of the mean $\kappa$ for the less steep and the more steep part of the scaling curves is shown in the parenthesis next to the value of the varied parameter, with ``-'' denoting a nonlinear scaling.
The solid gray lines in each panel have inclination of $\kappa=1$ and the dashed gray lines have inclination of $\kappa=0.25$.
}
\label{fig:scaling}
\end{figure*}

To study how the magnetic field strength scales with the local plasma density, we use the following approach. We examine only the field at the $xz$-midplane, which is the cross section of the middle part of the flux tube. 
Notice how the histogram of $B$ and $\rho$, at that plane and shown in  Fig.~\ref{fig:demonsrate_hist}, evolves in time. As the flux tube emerges (panels a to c), this histogram is shifted towards lower values of density and field strength. To track the overall change of the field strength with the local density, we make the histogram of all the values of $B$ and $\rho$, 
from $t=0$ until the end of the simulation
(an example is shown in Fig.~\ref{fig:scaling}a). 
We then plot the line that outlines the uppermost part of the histogram (black line). 
This line highlights how the maximum magnetic field strength scales with the local density (the undulations of this line are due to the snapshots frequency of the simulation: the higher the frequency, the smoother the line). 
We will refer to such lines as scaling curves.

We follow this process for the Table~\ref{tab:scaling} cases 1-15, which explore the $B_0$, $R$, $\alpha$, and $\lambda$ parameter space around the $B\dm=34$ ($B_0=10$~kG), $R\dm=5$ case of Fig.~\ref{fig:flux_B}. 
We show their scaling curves in Fig.~\ref{fig:scaling}b-f.
Notice that in most of the cases shown in Fig.~\ref{fig:scaling}, the scaling curves consist of two major parts: a less steep part ($\log\rho\in[-7,-6]$, i.e. $-7<\log\rho<-6$, where $\rho$ is in g~cm$^{-3}$) and a more steep part ($\log\rho\in[-5,-4.3]$). 
We identify the mean inclination ($\kappa$) of these parts by performing linear fits ($\log B=\kappa\log\rho+c$).
The value of $\kappa$ is shown in each panel, inside the parenthesis next to the value of the varied parameter.

As discussed in the Introduction, \citet{Cheung_etal2014} suggested that the scaling curve will change from a steeper ($\kappa=1$) to a less steep ($\kappa<1$) power law during the emergence within the solar convection zone. Fig.~\ref{fig:scaling}b-e shows a similar transition in our numerical experiments. 
However, Fig.~\ref{fig:scaling}f shows a number of simulations where the scaling curves are not linear at all. Why do some of the scaling curves develop power-laws while others behave non-linearly? How does the steep and less steep part of the scaling curves develop?
We first address these questions and then discuss how the variation of each flux tube parameter modifies the scaling curves and affects the emergence.

\subsubsection{Derivation of scaling laws and comparison with simulation}
\label{sec:scaling_laws}

In Appendix~\ref{app:appendix}, we derive forms to explain under which conditions the magnetic field strength scales with the local density as $B\propto\rho^\kappa$. Here we summarize these results.

First, we assume a velocity field with no shearing terms. We demand the magnetic field strength to be written as $B\propto\rho^\kappa$, where $\kappa$ is a constant. 
Combining the induction and continuity equation, we get Eq.~\ref{eq:kappa}, which we write again here:
\begin{align}
    B^2 \kappa = B_x^2 \kappa_x +  B_y^2 \kappa_y + B_z^2 \kappa_z, \nonumber
\end{align}
where $\kappa_x$, $\kappa_y$, $\kappa_z$ (Eq.~\ref{eq:kappa_x},~\ref{eq:kappa_y},~\ref{eq:kappa_z} respectively) are functions of the non-shear velocity gradients.
We then find solutions for $\kappa$ that satisfy the above equation. To do so, we focus on the following three cases.

The first case is when the magnetic field can be described with one component (e.g. $B_{i} \gg B_{j}, B_{k}$, where $i,\,j,\,k$ are the components of the field). Then, the magnetic field strength scales as $B\propto\rho^{\kappa_i}$. For instance, if $B_{x} \gg B_{y}, B_{z}$, then $\kappa=\kappa_x$, whereas  if $B_{y} \gg B_{x}, B_{z}$, then $\kappa=\kappa_y$.
The assumption of $\kappa$ being constant constrains $\kappa_i$ to be constant too. Therefore, the velocity field is constrained to be $\frac{\partial v_x}{\partial x} = \chi, \,
    \frac{\partial v_y}{\partial y} = \psi, \,
    \frac{\partial v_z}{\partial z} = \zeta,$
where $\chi,\,\psi,\,\zeta$ are constants.
Notice that the expression of $\kappa$ from \citet{Cheung_etal2010} (see Introduction) is a special case of the above. For their assumption that $\mathbf{B} = B \, \mathbf{\hat{x}}$, the field will scale with $\kappa_x$. Assuming the velocity gradients to be in the form of Eq.~\ref{eq:velocity_cheung}, then Eq.~\ref{eq:kappa_x} will give Eq.~\ref{eq:kappa_cheung}.

\begin{figure*}
\centering
\includegraphics[width=\textwidth]{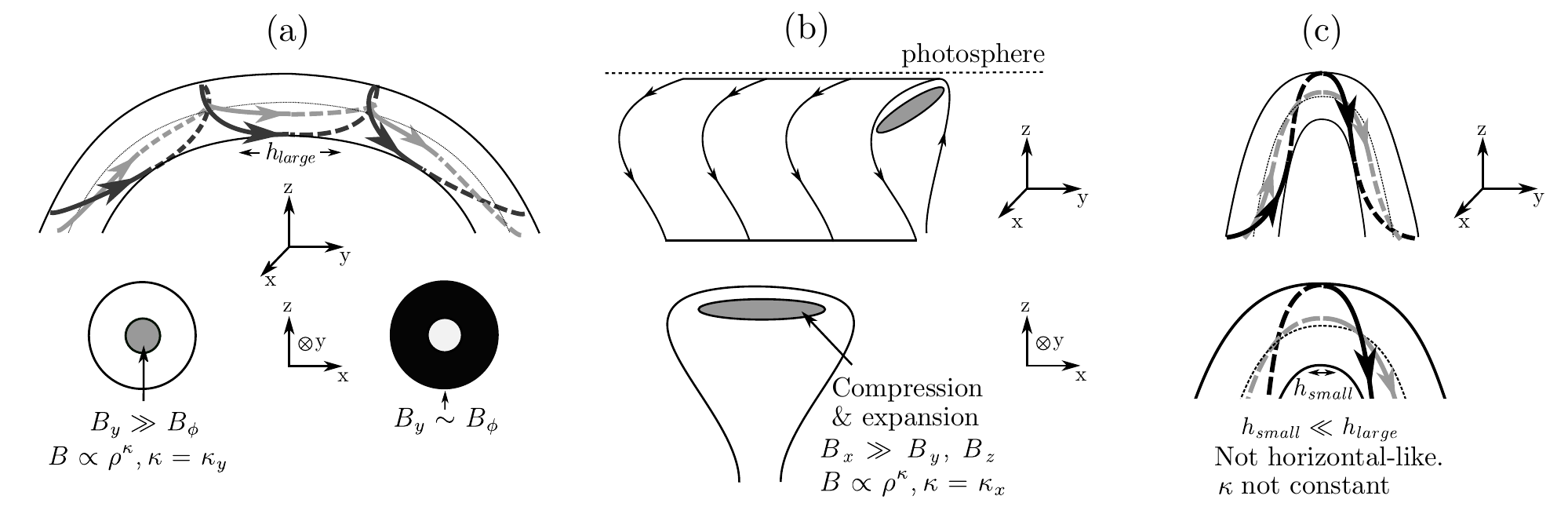}
\caption{
Cartoon-like illustration showing when and where $B\propto\rho^\kappa$ in an emerging flux tube. 
(a) Top: a flux tube with a horizontal-like apex. The thick grey (black) lines show twisted field lines close to (away from) the axis of the flux tube. 
Bottom: cross sections of the flux tube, with regions colored according to the color of the above field lines.
Bottom left: the region where the axial field is stronger than the poloidal field, and the corresponding scaling law. 
Bottom right: the region where the axial component is comparable to the poloidal component.
(b) Top: the horizontal expansion of the flux tube at the photosphere. The grey shaded region shows where the magnetic field strength increases due to compression.
Bottom: cross section of the flux tube, showing the compressed region and the corresponding scaling law. 
(c) Top: a flux tube with a toroidal-like shaped apex. The thick grey (black)lines show twisted field lines close to (away from) the axis of the flux tube. 
Bottom: blow-up of the apex of the tube.
}
\label{fig:cartoon}
\end{figure*}

The second case is when the magnetic field can be described with two components of the full magnetic field vector (e.g. $B_{i}, \, B_{j} \gg B_{k}$). Then, the magnetic field strength scales as $B\propto\rho^\kappa,\, \kappa=1- \frac{1}{2}\kappa_k$. For instance, if $B_{x}, \, B_{y} \gg B_{z}$, then the magnetic field strength will scale with $\kappa=1- \frac{1}{2}\kappa_z$ (Eq.~\ref{eq:kappa_2d}).
In this case, the constraint imposed on the velocity field will be stricter ($\frac{\partial v_x}{\partial x} = \frac{\partial v_y}{\partial y} = \chi, \quad
\frac{\partial v_z}{\partial z} = \zeta
$, where $\chi,\,\zeta$ are constants). This constraint guarantees that the magnetic field vector described by $B_{x}$ and $B_{y}$ does not change direction. As a result, in a high $\beta$ plasma (like a flux tube in the solar interior), such a velocity field will force the two-component field to behave as a one-component field. Therefore, in the field aligned coordinate system, this case is a special case of the first one.

The third case is when all three components of the magnetic field vector are important to describe the field. Then all $\kappa_x$, $\kappa_y$, $\kappa_z$ are equal. The velocity field constraint then becomes  $
\frac{\partial v_x}{\partial x} = \frac{\partial v_y}{\partial y} =\frac{\partial v_z}{\partial z} = \chi$, where $\chi$ is constant. 
So, the strength of the field will scale with the local density only if the field expands isotropically. In that case, $\kappa=\frac{2}{3}$ (Eq.~\ref{eq:kappa_3d}). In the field aligned system the three-component field behaves as a one-component field. Therefore this is also a special case of the first one.

From the above, we infer that in order to express the magnetic field strength as $B\propto\rho^\kappa$, the magnetic field needs to have one dominant direction. Otherwise, a power law cannot be derived in general.

In the above, we assumed that the shearing terms of the velocity field equal to zero, i.e. $\pder[v_i]{x_j}=0$, $i\neq j$.
Assuming constant $\kappa$ and a velocity field with non-zero shearing terms, we deduce that the magnetic field strength will scale with the local density with $\kappa = \frac{1}{B^2} \kappa_{ij} B_i B_j$ (Eq.~\ref{eq:k_shear}), where $\kappa_{ij}$ is a tensor describing the deformation of the velocity field, given by Eq.~\ref{eq:kij}. Because we assumed constant $\kappa$, the components of the tensor are required to be constant too. 
Note that the previous expressions derived for zero shear velocities are special cases of this general expression.

So far, we have assumed a strict power law between $B$ and $\rho$ (i.e. constant $\kappa$), which led to the constraint that the gradients of the velocity field components are constants. 
In general, the velocity gradients would be expected change during the emergence of a field.
Assuming a non constant $\kappa$, we derived that $\kappa$ can be described by Eq.~\ref{eq:kappa_equation}.
However, if $\kappa$ changes slowly both in space and time ($\lder[\kappa]{t}\approx0$), we get that $\kappa\approx\frac{1}{B^2} \kappa_{ij} B_i B_j$.
Therefore, the latter expression for $\kappa$ can describe the scaling of the $B$ with $\rho$, not only when $\kappa$ is constant, but also when $\kappa$ is changing slowly.

Our analysis suggests that when the velocity gradients change rapidly in space and/or time, or when the magnetic field cannot be adequately described by one component of the full magnetic field vector, $\kappa$ will not be constant. In that case, a power law between $B$ and $\rho$ should not be expected to occur.

We will now discuss where in our simulations we find conditions favouring the formation of power laws.
We focus on where the magnetic field has a dominant component during its emergence within the solar interior.
Notice that closer to the center of the flux tube, the poloidal component of the field becomes less significant than the axial one ($B_\phi/B_y$ decreases for smaller $r$, Eq.~\ref{eq:by},~\ref{eq:bphi}).
During the emergence of the flux tube, the shape of the field is crucial for the development of dominant field directions.
In our numerical experiments, the length scale of the buoyant part of the flux tube ($\lambda$) is the parameter that affects the shape of the apex of the emerging flux tube the most.
In Fig.~\ref{fig:cartoon}a, top,  we show a cartoon-like illustration of the upper part of a flux tube (oriented along the $y$-axis) with large $\lambda$. 
In this case, because of the large value of $\lambda$ the apex of the emerging tube is almost horizontal, oriented along the $y$-axis.
Close to the axis of the flux tube (grey line in Fig.~\ref{fig:cartoon}a top, grey shaded cross sectional region in Fig.~\ref{fig:cartoon}a bottom left) the axial ($B_y$) component of the field will be dominant across a length $h_{large}$. The magnetic field in this region is, therefore, expected to scale with $\kappa=\kappa_y$, if the velocity field is changing slowly. 
It is important to note that the magnitude of the magnetic field is stronger close to the flux tube axis (Eq.~\ref{eq:by}). 
Consequently, the region around the axis contains the bulk of the magnetic energy of the flux tube and, therefore, has the most important role in the transfer of that energy to the photosphere.
Away from the center, (thick black line in Fig.~\ref{fig:cartoon}a top, black shaded region Fig.~\ref{fig:cartoon}a bottom right) the poloidal component of the field becomes important. There, the field strength should not be expected to scale with the local density in general.

\begin{figure*}
\centering
\includegraphics[width=\textwidth]{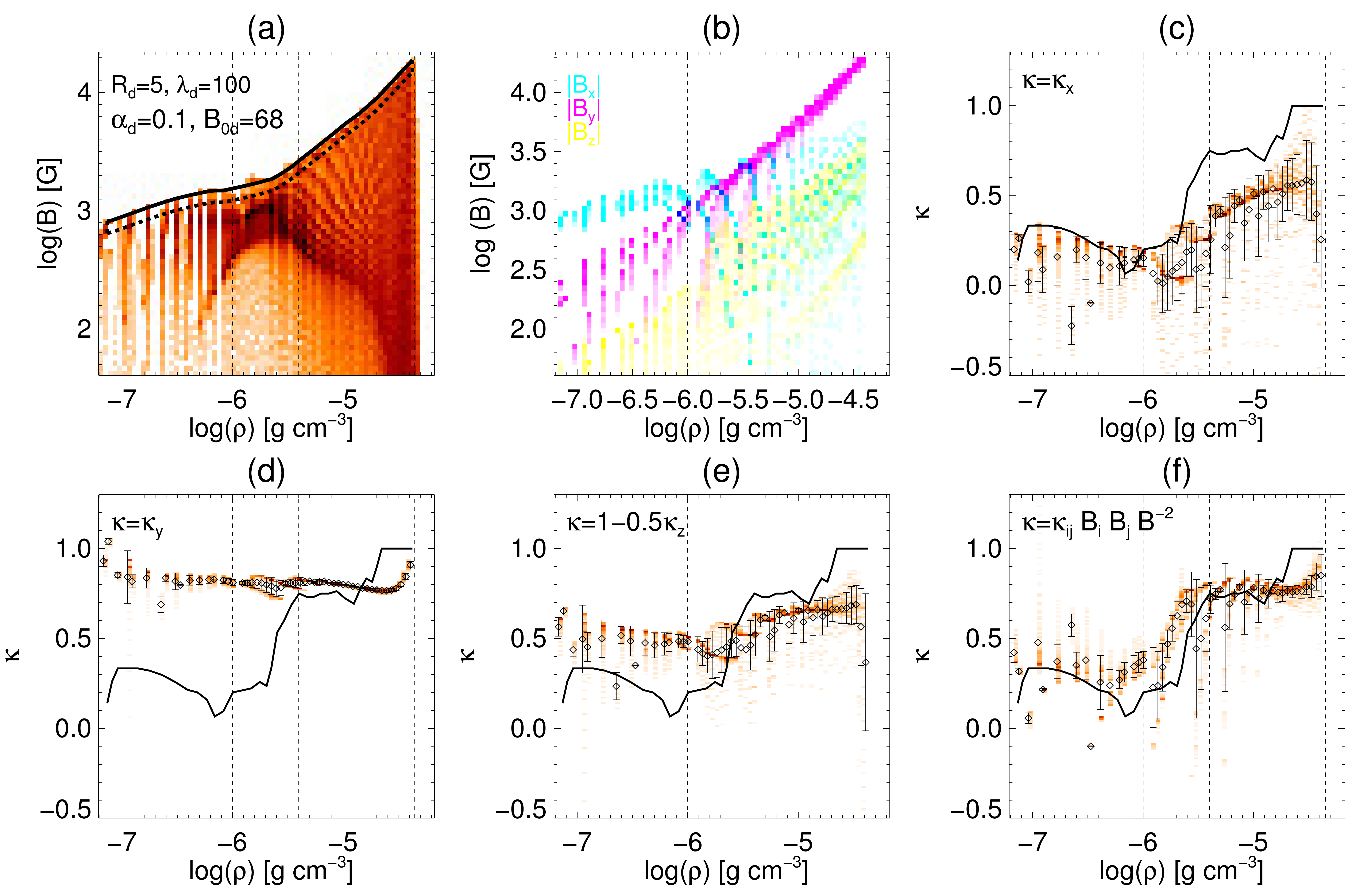}
\caption{
(a) Same as Fig.~\ref{fig:scaling}a, but for the $R\dm=5$, $\lambda\dm=100$, $\alpha\dm=0.1$ and $B_0\dm=68$ (case 4 Table~\ref{tab:scaling}). The solid line is the scaling curve. The dashed line is the scaling curved shifted by $\Delta\log B=0.2$. For the points between the solid and the dashed line, 
(b) shows the distribution of each component of the magnetic field vector (true color image, $B_x$ is blue, $B_y$ is magenta and $B_z$ is yellow), 
(c) shows the distribution of $\kappa=\kappa_x$, (d) shows the distribution of $\kappa=\kappa_y$, 
(e) shows the distribution of $\kappa=1-\frac{1}{2}\kappa_z$ and 
(f) shows the distribution of $\kappa=\kappa_{ij} B_i B_j /B^2$. 
The diamond symbols show the mean value of the distributions at each density bin, and the error bars show the standard deviation.
The solid black line in (c)-(f) is the derivative of the scaling curve (i.e. $\kappa$ measured from the scaling curve of panel (a)). The dashed vertical lines mark changes in the inclination of the scaling curve.
}
\label{fig:comparison_kappa}
\end{figure*}
During the emergence process, the tube expands and, hence, its radius increases. \citet{Parker_1974} showed that the radial expansion of a flux tube causes the poloidal component of the field to increase when the twist remains constant (i.e. for a tube oriented along the $y$-axis, $B_\phi/B_y$ increases). We do find $B_\phi/B_y$ increases, in agreement with Parker. Eventually, this effect would result in a decrease of the size of the region close to the axis that scales with $\kappa_y$ (shaded region in Fig.~\ref{fig:cartoon}a bottom left).

Fig.~\ref{fig:cartoon}b shows an illustration of the flux tube when its apex reaches the photosphere. The upper part of the tube (shaded region) undergoes compression and horizontal expansion.
If compressed enough, this region will develop locally a strong $B_x$ component. The $B_x$ component can eventually become much stronger than the local $B_y$ component of the field ( $B_x \gg B_y,\,B_z$). Then, the magnetic field strength inside the compressed region will scale with $\kappa=\kappa_x$ (and not with $\kappa=\kappa_y$, which was the scaling exponent during the rise of the flux tube deeper in the solar interior, where $B_y \gg B_x, \, B_z$ ).
Note that the axis of the flux tube might not be inside the compressed region. In fact, in our simulations the center of the tubes is found at lower heights.

Fig.~\ref{fig:cartoon}c (top) shows a flux tube that develops a highly bent apex when $\lambda$ is small. In this case, only a small segment of the apex (with a horizontal size $h_{small}$, Fig~\ref{fig:cartoon}c bottom) could be oriented parallel to the photosphere, adopting a horizontal-like configuration.
Also, due to the highly bent apex, plasma draining is expected to be more profound in this case, which could develop strong variations in the velocity gradients.
Thus, in this case, we should not expect $B$ and $\rho$ to scale with a power law, except if the tube undergoes significant compression at the photosphere and adopts a similar configuration to the case described in Fig.~\ref{fig:cartoon}b.


To use the above analysis towards studying the results of the simulations, we select an experiment with strong $B_0$ and large $\lambda$ (case with $B_0=20$~kG ($B_0\dm=68$), $\lambda\dm=100$, $\alpha\dm=0.1$ and $R\dm=5$) and find its scaling curve (Fig.~\ref{fig:comparison_kappa}a, solid line).
We shift the scaling curve by $\Delta\log B=0.1$ (dashed line) and we take into account all the points between the two curves. 
These are the points with very high $B$.
For these points, we plot the distributions of the absolute value of each individual component of the field ($|B_x|$ is the blue, $|B_y|$ is the magenta and $|B_z|$ is the yellow distribution, Fig.~\ref{fig:comparison_kappa}b). 
Then, we plot the distribution of $\kappa_x$, $\kappa_y$, $\kappa=1-\frac{1}{2}\kappa_z$ and $\kappa = \frac{1}{B^2} \kappa_{ij} B_i B_j$ (Eq.~\ref{eq:kappa_x}, \ref{eq:kappa_y}, \ref{eq:kappa_2d}, \ref{eq:k_shear}), calculated directly from the velocity field (oragne color, Fig.~\ref{fig:comparison_kappa}c, d, e, f).
The diamond symbols show the mean value of the distributions (i.e. mean value of $\kappa$) at each density bin, and the error bars show the standard deviation.
The black line in panels (c)-(f) show the derivative of the scaling curve (i.e. the $\kappa$ measured from the histogram). 

Fig.~\ref{fig:comparison_kappa}b shows that, for $\log\rho \in [-5.4,-4.35]$ (meaning $-5.4 < \log\rho < -4.35$, where $\rho$ is in ~g~cm$^{-3}$) or in terms of height $z\in [-18, -3 ]$~Mm), the strongest component of the magnetic field is $B_y$ (purple between second and third vertical line, in comparison to cyan and yellow). The large $\lambda$ ensures that the apex will be locally horizontal along a relatively large region, similar to Fig.~\ref{fig:cartoon}a 
Therefore, when $\log\rho \in [-5.4,-4.35]$, the steep power law segment of the scaling law should be described by $\kappa=\kappa_y$. Indeed, in Fig.~\ref{fig:comparison_kappa}d, the values of $\kappa_y$ measured from the velocity field (orange) and $\kappa$ measured from the gradient of the scaling curve (black line) are in a relatively good agreement.

The small buildup of $B_x$ and $B_z$ when $\log\rho \in [-5.4,~-4.9]$ (cyan and yellow, Fig.~\ref{fig:comparison_kappa}b) is due to the expansion of the flux tube (which increases $B_\phi/B_y$).
However, not many points have comparable $B_x$ and $B_y$. We should highlight that Fig.~\ref{fig:comparison_kappa}b is a true color image and the colors blend proportional to the value on the histogram. When comparable number of points have similar $B_x$ and $B_y$, cyan becomes purple. Therefore, when $\log\rho \in [-5.4,-4.35]$, the $B_\phi/B_y$ increase during the expansion of the flux tube is not significant, and it does not affect the steep power law. 

\begin{figure*}
\centering
\includegraphics[width=\textwidth]{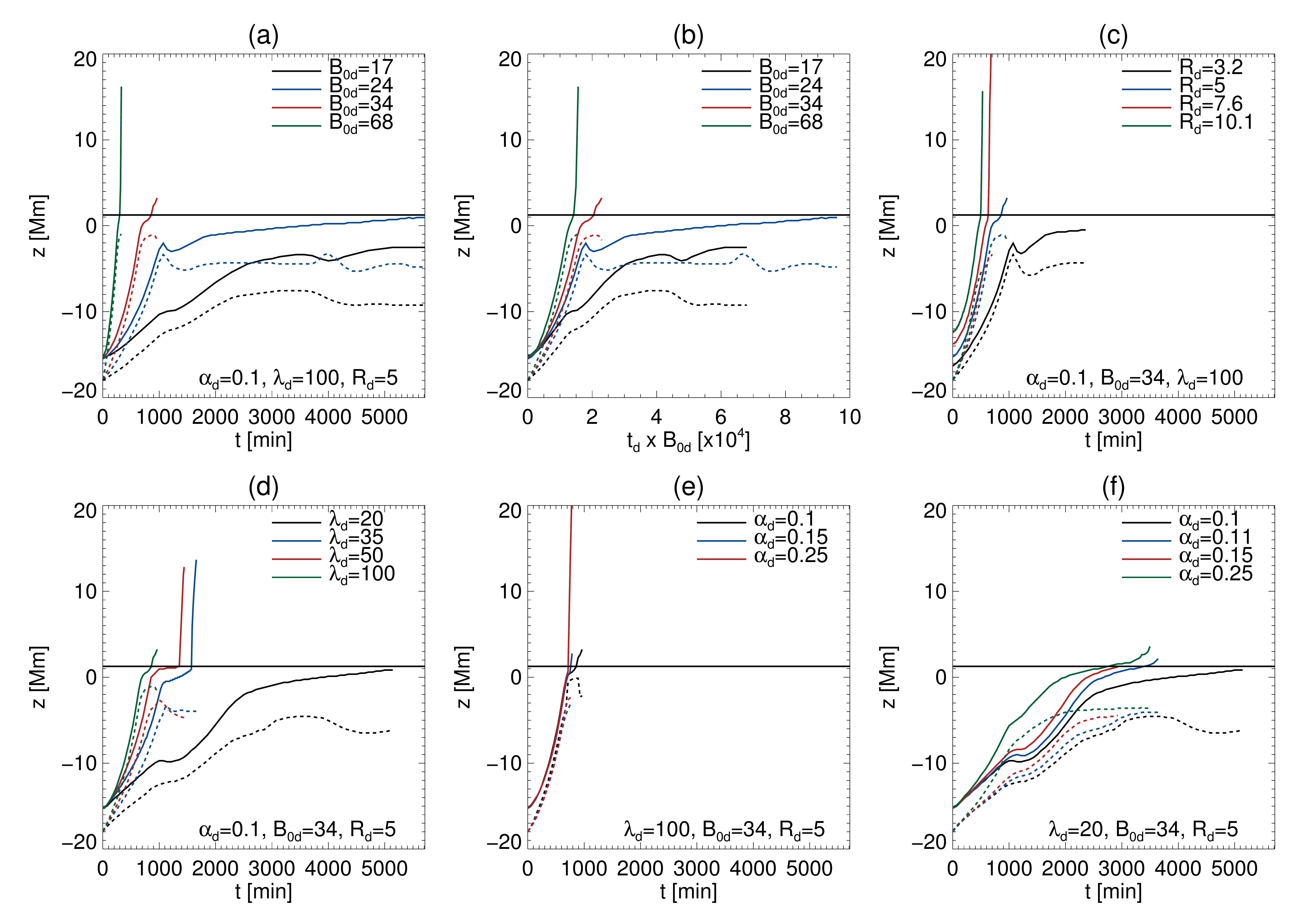}
\caption{
Height time profiles of the flux tube apex (solid) and center (dashed) of cases 1-15 of Table~\ref{tab:scaling}. 
Panel (a) shows cases with different $B_0$, (b) shows the same as (a) but the $x$-axis is scaled as $t\dm\times B_0\dm$, (c) shows cases with different $R$, (d) cases with different $\lambda$, (e) cases with different $\alpha$ for $\lambda\dm=100$ and (e) cases with different $\alpha$ for $\lambda\dm=20$.
}
\label{fig:h_t_profiles}
\end{figure*}

For values in the range $-6 < \log\rho < -5.4$ (or $z\in [-3, -1 ]$~Mm) the steepness of the scaling curve changes, revealing a transition to another regime with a different power law dependence between $B$ and $\rho$ (Fig.~\ref{fig:comparison_kappa}a, between first and second vertical line).
During that transition, $B_x$ increases and becomes comparable to $B_y$ (purple color).
The comparison between $\kappa$, deduced from the scaling curves, and the expression $\kappa=1-\frac{1}{2}\kappa_z$ (which is derived when both $B_x$ and $B_y$ are important) is in agreement at these depths (Fig.~\ref{fig:comparison_kappa}e, orange histogram and black line between first and second vertical lines).

For $\log\rho\lesssim-6$ or $z\gtrsim-1$~Mm, the apex of the flux tube is compressed significantly and $B_x$ becomes the strongest component of the magnetic field (Fig.~\ref{fig:comparison_kappa}b, cyan before first vertical line, in comparison to purple and yellow), as it is schematically illustrated Fig.~\ref{fig:cartoon}b. 
There, we find the less steep power law of the scaling curve. Since $B_x$ is significantly stronger than the other components of the field, the magnetic field strength is expected to scale with the local density raised to
the power $\kappa=\kappa_x$. Indeed, in Fig.~\ref{fig:comparison_kappa}c, the values of $\kappa_x$ measured from the velocity field and the $\kappa$ measured from the scaling curve are in good agreement.

Finally, we compare $\kappa$ measured from the scaling curve with the more general expression which includes velocity shear, $\kappa = \frac{1}{B^2} \kappa_{ij} B_i B_j$, calculated directly from the velocity field (Fig.~\ref{fig:comparison_kappa}f). We find that they are overall in agreement.

It is important to note that, for the derivation of the expressions of $\kappa$, we assumed that the velocity gradients are either constant or change slowly. For $\kappa_x$, $\kappa_y$ and $\kappa_z$, we also assumed zero shearing velocities. 
In the simulation, the velocity gradients are not changing slowly close to the photosphere. Also, the $\pder[v_y]{x}$ shear is significant when  $\log\rho\lesssim-5.4$.
However, the expressions of $\kappa$ shown in Fig.~\ref{fig:comparison_kappa}c, d, e, f, are in agreement with the values measured from the scaling curve. 
Therefore, we conclude that the most important parameter for the development of the power laws is a strong locally horizontal field across a large region, and not the strict velocity field constraints.
However, we expect that for significant variations of the velocity gradients, which can perturb the direction of the magnetic field, it is not possible to form at power law.

The effects that the resolution, resistivity and viscosity have on the scaling curve of the studied case is discussed in Appendix~\ref{app:appendixB}.

\subsection{Height-time profiles and scaling curves of parametric study}
\label{sec:parametric}
In the following, we will study how the initial parameters of the flux tube (e.g. $B_0$, $R$, $\lambda$, $\alpha$) affect the emergence to the photosphere and above.
For that, we focus on the cases 1-15 of Table~\ref{tab:scaling}, which explore the parameter space around the $B_0=10$~kG and $R\dm=5$ point of Fig.~\ref{fig:flux_B}.
We will study the emergence, focusing on the height-time profiles and the scaling curves of the emerging fields.

To plot the height-time profiles, we follow the rising motion of two points of the emerging flux tubes. The first one is the center of the flux tube, which is the point where $B_y$ is maximum and $B_x$ changes sign, along the $z$-axis at the center of the numerical box. The second one is the apex of the rising flux tube, which we consider to be the uppermost point along the $z$-axis at the center of the numerical domain, where $B>0.001 B_0$.
The profiles are plotted in Fig.~\ref{fig:h_t_profiles} with solid (apex) and dashed (center) lines.

\subsubsection{Variation of magnetic field strength}
\label{sec:field_strength}

Firstly, we focus on the dynamics of the emerging flux tube when the magnetic field strength is varied and the other parameters are kept constant. 
We select $B_0=5,\, 7.2, \, 10,\, 20$~kG ($B_0\dm= 17, \, 24,\, 34,\, 68$) and $\alpha\dm=0.1$, $\lambda\dm=100$, $R\dm=5$ (Table~\ref{tab:scaling}, cases 1-4).

The height-time profiles are shown in Fig.~\ref{fig:h_t_profiles}a.
It is clear, that the stronger the field strength the faster the flux tube will rise inside the solar interior. 
Notice that the $B_0\dm=68$ case emerges above the photosphere almost immediately. 
In comparison, the $B_0\dm= 34$ case exhibits a phase of deceleration before it emerges above the photosphere (during which the magnetic field at the apex is locally compressed). 
This is consistent with the results reported in previous studies \citep[e.g.][]{Fan_2001,Archontis_etal2004, Toriumi_etal2013}. 
For lower $B_0$, the buoyancy of the flux tubes decreases and, thus, their center reach lower heights in the convection zone.
The cases where $B_0\dm=17, \,24$ ``failed'' to emerge. 

If we scale the time as $t \times B_0$ (Fig.~\ref{fig:h_t_profiles}b), we find that the height-time profiles are ``clustering'' closer together indicating self-similar behaviour \citep{Murray_etal2006, Sturrock_etal2016}.
Still, the ``clustering'' is not as ``compact'' as in the previous studies.
In our simulations, the flux tubes emerge from much deeper down in the solar interior. Thus, the downwards tension force becomes higher, reducing the upwards buoyancy force.

Fig.~\ref{fig:scaling}b shows the scaling of $B$ with $\rho$. We focus on the steeper part of the scaling curves. Increasing $B_0$ decreases the value of $\kappa$ (from $\kappa$=2.2 in the $B_0\dm=17$ case to $\kappa$=0.76 in the $B_0\dm=68$ case). 
Therefore, flux tubes with higher $B_0$ emerge more efficiently. 
In the ``failed'' emergence cases, the central part of the emerging fields reach moderate heights within the convection zone (around -9~Mm and -5~Mm for $B_0\dm=17$ and $B_0\dm=24$ respectively, Fig.~\ref{fig:h_t_profiles}a). The apexes  move slowly upwards, but never emerge through the photosphere. 
Because of the lower $B_0$, these flux tubes do not undergo a 3D full expansion but mainly experience a vertical stretching in the following manner. The lower segments of the buoyant part of the flux tubes remain almost anchored at the initial depth. The rest of the tube emerges slowly, causing the vertical stretching.
This stretching leads to faster decrease of the axial field strength and as a result a higher $\kappa$ ($\kappa>1)$.

The transition to the less steep part of the scaling curves occurs when the flux tubes are close to the photosphere (around -4~Mm or $\log(\rho)=-5.3$).
There, the scaling curve transitions from scaling with $\kappa=\kappa_y$ to scaling with $\kappa=\kappa_x$.
The ``failed'' emergence cases with $B_0\dm=17$ and $B_0\dm=24$ do not experience significant compression and, therefore, do not develop the less steep slope.

\subsubsection{Variation of radius}
\label{sec:radius}

Next, we focus on the dynamics of the emerging flux tubes when their radius is varied.
We select $R\dm= 3.2, \, 5,\, 7.6,\, 10.1$, and $\alpha\dm=0.1$, $\lambda\dm=100$, $B_0=10$~kG ($B_0\dm=34$) (Table~\ref{tab:scaling}, cases 3, 5, 6, 7). 
The height-time profiles of these cases are shown in Fig.~\ref{fig:h_t_profiles}c. 
Notice that the larger the radius the faster and higher the flux tube's apex and center will rise. 
At $t=0$, all the $B_0=10$~kG flux tubes are equally buoyant (buoyancy $\propto B^2$) at their centers (where $B=B_0$). 
However, they are not equally buoyant away from their centers, as $B_{R_{large}}>B_{R_{small}}$ when $r>0$ (see Eq.~\ref{eq:by},~\ref{eq:bphi}). 
Thus, a larger radius tube will be more buoyant across its whole cross section.

As discussed in Sec.~\ref{sec:fluxes}, the magnetic pressure of smaller radius flux tubes will decrease faster. 
This can be seen in the steeper part of the scaling curves in Fig.~\ref{fig:scaling}c. 
Notice that $\kappa$ decreases as $R\dm$ increases, both at the steeper ($\kappa=2.3,\,0.98,\,0.86,\,0.77$ for the $R\dm=3.2,\,5,\, 7.6,\, 10.1$ cases respectively) and at the less steep part of the scaling curve ($\kappa=0.23,\,0.16,\,0.15$ for the $R\dm=5,\, 7.6,\, 10.1$ cases). 
Overall, higher $R$ leads to more efficient emergence. 
This is also reflected in the time needed for the flux tube to emergence above the photosphere. For instance, in Fig.~\ref{fig:h_t_profiles}, the $R\dm=10.1$ (green) flux tube emerges almost directly in comparison to the $R\dm=5$ (blue) case.
Therefore, the radius of the tube is an important parameter affecting the dynamics of the emergence.

Notice that the point where the scaling curves transitions from the steep to the less steep power law is different for each case. It can be traced approximately at $\log(\rho)=-5.3,\,-5,\,-4.7$, where $H_p\approx2R$ (local pressure scale of $H_p\dm=9,\,14,\,22$) for the $R\dm=5,\, 7.6,\, 10.1$ cases respectively).

\subsubsection{Variation of $\lambda$}
\label{sec:lambda}

We now focus on the dynamics of the emerging flux tubes when the length of their buoyant part is varied. We select $\lambda\dm=20,\,35,\,50,\,100$ and $\alpha\dm=0.1$, $B_0\dm=34$ and $R\dm=5$ (Table~\ref{tab:scaling}, cases 3, 8, 9, 10). 

The buoyant part of the flux tube becomes more bent for smaller $\lambda$,  resulting  to higher downward magnetic tension at its apex. Due to that, smaller $\lambda$ flux tubes emerge slower (Fig.~\ref{fig:h_t_profiles}d) \citep[e.g.][]{Schuessler_1979,Longcope_etal1996,Moreno-Insertis_etal1996}.
The $\lambda\dm=35,\,50,\,100$ results are consistent with the results of previous studies \citep[e.g.][]{Fan_2001,Syntelis_etal2015}.

However, the $\lambda\dm=20$ case (Fig.~\ref{fig:h_t_profiles}d , black line) behaves differently. This is a case of a ``failed'' emergence. Initially, the flux tube rises for a time period of about $t=1000$~minutes (black solid and dashed line). Then, the emerging flux system enters a short phase of deceleration (i.e. from $t=1000$~minutes until $t=1400$~minutes), during which the downward tension force of the envelope field lines becomes comparable to the magnetic pressure force. 
At the same time, plasma draining from the apex of the tube towards its flanks, becomes more efficient due to the highly curved shape of the flux tube. 
The draining makes the flanks significantly heavier than the surrounding material. Thus, while the apex continues to emerge, the flanks start to submerge. 
The submergence modifies the geometrical shape of the emerging field further, making the apex of the flux tube even more curved and, therefore, further enhancing the plasma draining.
Eventually (after $t=1400$~minutes), the middle part of the flux tube loses enough mass to become buoyant again, and, hence, continues to rise and to expand. 
This complicated process affects the overall horizontal and vertical expansion the flux tube, resulting in a reduced magnetic field strength. 
Thus, when the flux tube reaches the photosphere, it carries very high $\beta$ plasma. Furthermore, the compression rate of the field below the photosphere is very low. As a result, the field fails to emerge.

The biggest difference between $\lambda=20$ and higher $\lambda$ cases is found at the scaling curves (Fig.~\ref{fig:scaling}d). The $\lambda=20$ case does not scale with a power law. Notice also that the magnetic field strength, $B$ is significantly reduced as the field rises.
Interestingly, the variation of $\lambda$ from 35 to 100 does not effect dramatically the scaling curves. 

\subsubsection{Variation of $\alpha$}
\label{sec:alpha}

We now focus on the dynamics of the emerging flux tubes when the twist is varied.
We showed that small $\lambda$ affects significantly the plasma draining along the field lines. 
The twist is a parameter that affects the efficiency of the draining, as higher twisted field lines has \lq dips\rq  that can trap dense plasma.
We will study the effects of the variation of twist by using both large and small $\lambda$ to capture the effect of the twist on the draining along the field lines.
\begin{figure*}
\centering
\includegraphics[width=\textwidth]{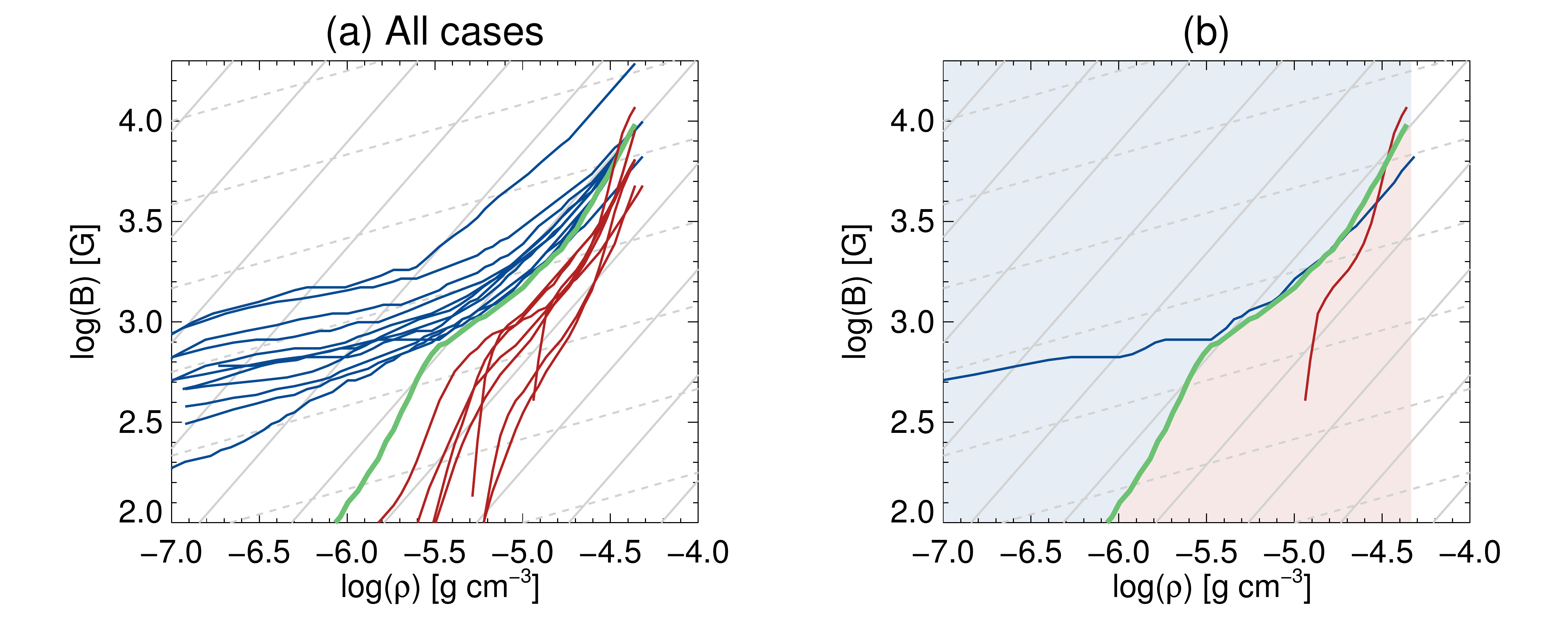}
\caption{
(a) This panel shows the scaling curves of all the Table~\ref{tab:scaling} cases. 
Blue lines show cases that ``successfully'' emerged and red lines show cases that ``failed'' to emerge. The green line show the ``failed'' emergence of Case 8, Table~\ref{tab:scaling}, which separates most of the ``successful'' and ``failed'' emergence cases. 
(b) This panel shows the green line of panel (a). We color the region above the green line with blue color (inside which most ``successful'' emergence cases are located) and the region below the green line with red color (inside which most ``failed'' emergence cases are located).
The blue line shows the scaling of $B$ with $\rho$ of the ``successful'' emergence of Case 20, Table~\ref{tab:scaling}. 
The red line shows the scaling of $B$ with $\rho$ of the ``failed'' emergence of Case 17. 
}
\label{fig:scaling_all}
\end{figure*}

First, we select values of $\alpha\dm=0.1,\,0.15,\,0.25$ and $\lambda\dm=100$, $B_0\dm=34$, $R\dm=5$  (Table~\ref{tab:scaling}, cases 3, 11, 12).
The larger, twist flux tubes emerge slightly faster (Fig.~\ref{fig:h_t_profiles}e).
This is consistent with previous studies \citep[e.g.][]{Murray_etal2006}. 
Notice also that their scaling curves behave similarly, deeper in the convection zone (steeper slopes in Fig.~\ref{tab:scaling}e). 
Closer to the photosphere, the higher the twist, the smaller the value of $\kappa$ ($\kappa=0.23,\, 0.13,\, 0.11$ for $\alpha\dm=0.1,\,0.15,\,0.25$).
This is expected as i) the radial magnetic tension from the twist keeps the flux tube more coherent, bringing stronger field below the photosphere and ii) higher twist flux tubes have stronger poloidal field component, which is further enhanced during the compression below the photosphere. So, overall, the flux tubes with higher twist emerge more efficiently.

Now, we select cases with a smaller $\lambda\dm=20$ and $\alpha\dm=0.1,\,0.11,\,0.15,\,0.25$, $B_0\dm=34$, $R\dm=5$ (Table~\ref{tab:scaling}, cases 8, 13, 14, 15).
In the low $\lambda$ cases we find some unexpected results.

The cases with $\alpha\dm=0.1,\,0.11,\,0.15$, at $t\approx1000$~minutes,
stop rising for a small time period  (Fig.~\ref{fig:h_t_profiles}f). Then they start rising again until they become decelerated by the photosphere. This is similar to the $\lambda=20$ case of Sec.~\ref{sec:lambda}. The net effect of this motion is enhanced plasma draining, leading to a complicated horizontal and vertical expansion.
However, the $\alpha\dm=0.25$ flux tube behaves differently (green lines).
There, the higher twist prohibits the enhanced draining that occurs in the lower $\alpha\dm$ cases. 
This flux tube emerges without the complicated horizontal and vertical expansion that is present in the lower twist cases.
As a result, its internal magnetic pressure is reduced less during its emergence.
However, the high downwards magnetic tension and the lack of draining eventually reduces the rate of emergence of the $\alpha\dm=0.25$ case ($t=1200-2500$~min, green line). 

For the cases with the enhanced draining ($\alpha\dm=0.1,\,0.11,\,0.15$), increasing $\alpha$ led to more efficient emergence (Fig.~\ref{fig:scaling}f). We do not find power laws deep in the convection zone for these cases. 
A less steep linear power law appears only for $\alpha\dm=0.11,\,0.15$, when they compress below the photosphere. 
However, for $\alpha\dm=0.25$, due to the deceleration of the flux tube, the compression below the photosphere is less. Thus, less steep part of the scaling curve has higher $\kappa$ value than the values for the less twisted cases.
Therefore, we find that for $\lambda\dm=20$, the higher twisted flux tube emerges less efficiently than the less twisted cases.

\subsubsection{All cases}
\label{sec:all_cases}

We now plot all the scaling curves (Table~\ref{tab:scaling}, cases 1-15) in Fig.~\ref{fig:scaling_all}a. We also plot some additional cases that mostly describe ``failed'' emergence (cases 16-21). The blue lines are the cases that ``successfully'' emerge above the photosphere (non-asterisk cases in Table~\ref{tab:scaling}) and the red lines are the cases that ``fail'' to emerge above the photosphere (asterisk cases). 
Notice that there is a clear separation and clustering of the blue and the red lines. The green line is the scaling curve of the ``failed'' emergence of case 8 (discussed in Sec.~\ref{sec:lambda}), which acts as a ``border-line'' between the bulk of the ``successfully'' emerged cases and the ones that ``failed'' to emerge.

In Fig.~\ref{fig:scaling_all}b, we plot again the ``border-line'' case (green line). We color the region above that line with blue and below with red. 
We noticed that if the left-most part of flux tube's scaling curve is located inside the blue region, then parts of the flux tube will eventually emerge above the photosphere (case 20, blue line, Fig.~\ref{fig:scaling_all}b). 
If it ends inside the red region, it will eventually fail to emerge (case 17, red line). 
Using the above comments and how $\kappa$ behaves when varying $B_0$, $R$, $\alpha$, we were able to estimate flux tube parameters needed for a ``successful'' or ``failed'' emergence, and roughly estimate the value of the magnetic field below the photosphere.

An interesting result is the following. In Fig.~\ref{fig:scaling_all}a, most of the blue lines originate from the same point, as they initially have a $B_0=10$~kG field. However, the photospheric field is much different. Therefore, the efficiency of the emergence (ratio of maximum photospheric field strength over $B_0$) is different. For instance, case 15 has an efficiency of 0.02, while case 7 has an efficiency of 0.1. The rest of the ``successful' emergence cases starting with $B_0=10$~kG have intermediate values of efficiency. 
This difference in the efficiency, is due to effects of the geometry of the field (twist, radius, curvature) on the emergence.
Fig.~\ref{fig:scaling_all}b is another example of the effect of the geometry on the emergence. Case 17 (red line) is a case with $B_0\dm=44$ ($B_0=13200$~G), that fails to emerge because of its very small $\lambda\dm=10$. However, case 20 (blue line), which has similar flux to case 17, but almost half the magnetic field strength ($B_0\dm=24$ ($B_0=7200$~G)), emerges ``successfully'' due to the larger radius and the larger $\lambda$.

The physical meaning of the above is that, in order for a flux tube to emerge above the photosphere, it must bring with it the necessary amount of magnetic field strength and flux.
If its geometry and twist are not favouring the efficient emergence of this field, then even an initially strong field will fail to emerge. On the other hand, weaker fields can emerge above the photosphere if their geometry results to a more efficient emergence. Our ``border-line'' case is a numerically derived limit that separates the two states.

\section{Summary and Discussion}
\label{sec:conclusions}

In this work we studied the emergence of flux tubes from 18~Mm below the photosphere, using 3D MHD numerical simulations. We performed a detailed parametric study on: (i) the magnetic field strength; (ii) the twist; (iii) the radius and (iv) the length of the buoyant part of a flux tube.
Initially, we varied the radius and the magnetic field strength (while keeping the twist and the length of the buoyant part constant), to study whether the initial amount of sub-photospheric magnetic flux is a good indicator for ``successful'' emergence (Fig.~\ref{fig:flux_B}). 
Then, we focused on the scaling of the maximum magnetic field strength with the local density. 
We identified the curve that describes the maximum $B$ as a function of $\rho^\kappa$ (scaling curve). 
The scaling curve had a part with steeper slope (larger $\kappa$, where $\kappa$ is the power of the density such that $B\propto\rho^\kappa$) and this developed in the deeper part of the solar convection zone. 
Close to the photosphere, $B$ scales with $\rho^\kappa$ with a smaller $\kappa$. However, in a few cases, the curves did not follow such a power laws.
We identified under which conditions the scaling curve can form a power law, and derived expressions for $\kappa$ that describe approximately the scaling.
Finally, we studied the scaling curves and the height-time profiles for a number of different initial conditions (Table~\ref{tab:scaling}) by keeping constant three of the $B_0,\,R,\,\alpha,\,\lambda$ and varying the remaining variable (Fig.~\ref{fig:scaling}, \ref{fig:h_t_profiles}). 

Our results are summarized as follows:
\begin{enumerate}[itemsep=0mm]
    \item Magnetic flux alone is not sufficient to estimate whether the magnetic field will emerge, especially below $10^{21}$~Mx.
    \item $B$ scales as $\rho^\kappa$ when the magnetic field has one dominant direction (the apex of the emerging flux tube is locally horizontal along a large enough segment) and the spatial/temporal changes of the velocity gradients and shear are not significant. In its most general form, a constant $\kappa$ can be described by Eq.~\ref{eq:k_shear}.
    \item The steeper part of the scaling curves develops when the flux tube apex is horizontal-like and is located deeper in the solar interior (similar to Fig.~\ref{fig:cartoon}a).
    The less steep part of the scaling curves develops due to the compression of the flux tube just below the photosphere (similar to Fig.~\ref{fig:cartoon}b).
    The transition from the less to the more steep part of the scaling curve occurs approximately when the characteristic radial size of the emerging tube is similar to the local pressure scale height ($2R\approx H_p$ in our case). Some parameters (like twist) can affect this characteristic length as they affect the rate of the flux tube expansion.
    For flux tubes whose apex is not horizontal-like, the field strength does not scale with the local density deeper in the solar interior (similar to Fig.~\ref{fig:cartoon}c). However, a power law can be developed below the photosphere if such a flux tube compresses significantly.
    \item The magnetic field is transferred upwards more efficiently when $B_0$ or $R$ is increased.
    In most cases, this applies also for the twist.
    \item A highly curved flux tube (small $\lambda$) with low twist emerges less efficiently in comparison to a lower curvature flux tube (large $\lambda$) with similar twist.
    \item In a highly curved flux tube, increasing the twist increases the efficiency of the emergence to a certain extent. Eventually, the higher twist obstructs the plasma draining by maintaining a local dip in the magnetic field, the flux tube remains heavy and the efficiency of emergence is reduced.
    As a result, a higher twisted flux tube can eventually bring less magnetic field closer to the photosphere in comparison to a less twisted one.
    \item The combined effect of all the above (Fig.~\ref{fig:scaling_all}a) shows that the efficiency with which the magnetic field is brought upwards is a significant aspect of the emergence of buoyant magnetic fields in the solar interior
    For instance, high-$B_0$ (weak-$B_0$) fields may fail (succeed) to emerge to the photosphere, depending on their geometrical properties. 
\end{enumerate}

Based on our results, it is clear that there is neither a specific $\kappa$ for which $B\propto \rho^\kappa$ everywhere in the solar interior nor a specific $\kappa$ that describes the field close to the photosphere. 

Deep in the solar interior, \citet{Pinto_etal2013} found in their dynamo simulation that $\kappa\approx1$. They showed that the poloidal expansion dominated over the axial expansion. This is in agreement with our analysis. Assuming a strong axial field oriented along the $x$-axis, the field would scale with $\kappa_x$. Then, for negligible axial expansion ($\pder[v_x]{x}\approx0$), from Eq.~\ref{eq:kappa_x} we derive that $\kappa\approx1$.
If the axial expansion is not negligible in comparison to the poloidal one, the value of $\kappa$ can be different. 

\citet{Cheung_etal2010} studied $\kappa$ in the case of the emergence of a highly twisted toroidal flux tube inside a convective layer and found $\kappa=0.5$. 
We consistently find lower values than that, meaning that in our simulations the magnetic field is transferred more efficiently upwards.
It is possible that this is due to the lack of a fully developed convective envelope in our or their simulations. Convective motions should deform to an extent the flux tubes, and reduce the efficiency of emergence.
Thus, the effect that convective motions have on emerging flux tubes is very important for the study of the scaling of $B$ with $\rho$. 

Such effects cannot be easily estimated.
However, the comparison between the buoyancy force and the drag force has been proposed as a measure for identifying whether convective motions will have a destructive effect on a flux tube or not.
\citet{Moreno-Insertis_1983, Fan_etal2003, Cheung_etal2007} showed that the flux tube will not be fragmented by the convective motions if its magnetic field strength is:
\begin{align}
    B\gtrsim  \sqrt{\frac{H_p}{R}} \, B_{eq}\,,
    \label{eq:equipartition}
\end{align}
where $B_{eq}$ is the equipartition value of the magnetic field strength with the local kinetic energy density ( $B_{eq} = \sqrt{\mu \rho u_{downflow}}$, where $ u_{downflow}$ is the local velocity of downdrafts). To estimate $B_{eq}$, we need the velocities of the local vertical flows.

In helioseismology, vertical velocities are calculated by averaging data across large regions \citep[e.g.][]{Komm_etal2004,Komm_etal2011}. 
Therefore, the local fast upflows and downflows are smoothed and such vertical velocities estimate the mean value across these large regions.
Also, comparisons between models and helioseismology methods have posed questions about the accuracy of vertical velocity measurements below certain depths \citep{Zhao_etal2010}.
As a result, we cannot use vertical velocities from helioseismology to estimate $B_{eq}$ in Eq.~\ref{eq:equipartition}. 
To estimate $B_{eq}$, we could assume some values for the vertical velocities. For instance, we can assume that the vertical velocities are of the order of the horizontal velocities derived from helioseismology \citep[e.g.][]{Greer_etal2015}.
Another approach could be to use the vertical velocities at different depths given from models \citep{Stein_etal2011}.
Using either the vertical velocity root mean square from \citet[e.g.][]{Stein_etal2011}, or the horizontal velocity root mean square from \citet{Greer_etal2015} we find that our selected $B_0$ values satisfy Eq.~\ref{eq:equipartition}. 
So, our flux tubes would not be fragmented from the downdrafts, at least deeper in the interior.
However, as the flux tubes expand closer to the photosphere, we expect that the convective motions will deform these flux tubes, reducing the magnetic field strength. We would expect also to find the ``opposite'' of the deformation. In a fully developed convective layer weaker fields can intensify locally due to ``convective intensification'' \citep[e.g.][]{Parker_1978,Spruit_1979}. 
The actual degree of deformation and intesification, their effect on $\kappa$ and whether they could significantly impact the emergence of magnetic elements above the photosphere, is unknown. 
To estimate these effects requires 3D compressive simulations with fully developed convection zones. 

Note that, in this work, we do not aim to identify conditions where flux tubes will form an active region (of any size). Our aim is to study the scaling, and identify cases where the field emerges above the photosphere, even if the photosheric magnetic field strength is small.

In most 3D flux emergence simulations, the flux tube is initially located close to the photosphere, around $-5$~Mm to $-1$~Mm \citep[e.g.][]{Fan_2001,Magara_etal2001,Manchester_etal2004, Archontis_etal2004,Murray_etal2006,
Hood_etal2009,MacTaggart_etal2009b,Fan_2009,Moreno-Insertis_etal2013,Toriumi_etal2013, Leake_etal2013,Fang_etal2014,Lee_etal2015,Takasao_etal2015, Syntelis_etal2015,Syntelis_etal2017}.
For instance, in the parametric study of \citet{Murray_etal2006}, the flux tube is placed at $-1.7$~Mm, whereas in our simulation the flux tube is placed at $-18$~Mm. 
We found that the previous results are consistent with the results of flux tubes placed deeper in the interior. However, our work shows that additional effects are also important during the emergence of flux tubes from deeper in the interior, associated mostly with the plasma draining along the field lines. 
\citet{Toriumi_etal2013} performed 3D simulations of flux tubes placed at -20~Mm. They did not find the effects on the plasma draining that we identified in our simulations when varying $\lambda$ and the twist. This is 
probably because they did not explore the same parameter space of low $\lambda$ and twist. For flux tubes similar to theirs, our results are in agreement. On the other hand, they showed that for higher values of $\lambda$ (e.g. $\lambda\dm=400$) than the ones we used, the flux emerges slightly slower compared to lower $\lambda$ cases (e.g. $\lambda\dm=100$). They attributed this behaviour to very slow plasma draining. We do not find a similar behaviour in our simulations, but it is possible that the further increase of $\lambda$ could lead to similar results.

Notice that in Fig.~\ref{fig:scaling_all}a, the vast majority of the ``successfully'' emerged cases (blue lines) start with the same $B_0$, and differ in flux, twist and the $\lambda$. Just below the photosphere, though, the magnetic field strength ranges from $200-1000$~G. 
Therefore, the magnitude of the photospheric magnetic field does not contain sufficient information to infer the magnetic field strength of the initial flux tube.
To estimate $B_0$, information about the radius and the shape of the flux tube are needed. This information, along with some estimate of the subphotospheric velocity vector, can assist in estimating the value of $\kappa$ close to the photosphere and the depth where the scaling curve changes behaviour. Hence, this could be used to estimate the magnetic field strength deeper in the interior. 
For such a calculation, further work is needed in many aspects. For instance, using 3D flux emergence models, it is important to identify whether the photospheric values of twist and the length scale of the emerged field can be correlated with the corresponding subphotospheric values. If no such relation exists (similar to our result for $B_0$ and photospheric $B$), then using the photospheric values of twist, $B$, and the size of an active region would provide little information about the conditions below the photosphere.
To understand the nature of the magnetic fields below the photosphere, information about the sub-photospheric magnetic field strength and the sizes of the typical emerging structures is required. Such parameters are essential to further develop our understanding of solar flux emergence and to pose constrains on numerical models.

\acknowledgments
This project has received funding from the Science and Technology Facilities Council (UK) through the consolidated grant ST/N000609/1.
The authors acknowledge support by the Royal Society.
This work was supported by computational time granted from the Greek Research \& Technology Network (GRNET) in the National HPC facility - ARIS. 
\appendix
\section{Derivation of scaling laws}
\label{app:appendix}

\subsection{Velocity field without shearing terms}
\label{app:without_shear}

Assuming a velocity field with no shearing $( \frac{\partial v_i}{\partial x_j} = 0, \quad i\neq j )$, the components of the ideal induction equation (Eq.~\ref{eq:induction}) can be written as:
\begin{subequations}
\begin{align}
    \frac{D B_x}{Dt} &= - B_x \kappa_x \mathbf{\nabla}\cdot\mathbf{v} \label{eq:bx_induction}\\
    \frac{D B_y}{Dt} &= - B_y \kappa_y \mathbf{\nabla}\cdot\mathbf{v} \label{eq:by_induction}\\
    \frac{D B_z}{Dt} &= - B_z \kappa_z \mathbf{\nabla}\cdot\mathbf{v} \label{eq:bz_induction},
\end{align}
\end{subequations}
where we define
\begin{subequations}
\begin{align}
     \kappa_x = 1 - \frac{1}{ \mathbf{\nabla}\cdot\mathbf{v}} \frac{\partial v_x}{\partial x} \label{eq:kappa_x}\\
     \kappa_y = 1 - \frac{1}{ \mathbf{\nabla}\cdot\mathbf{v}} \frac{\partial v_y}{\partial y} \label{eq:kappa_y}\\
     \kappa_z = 1 - \frac{1}{ \mathbf{\nabla}\cdot\mathbf{v}} \frac{\partial v_z}{\partial z}. \label{eq:kappa_z}
\end{align}
\end{subequations}
Combining Eq.~\ref{eq:bx_induction},~\ref{eq:by_induction},~\ref{eq:bz_induction} we get:
\begin{align}
    \lder[B^2]{t} & =
    - 2 ( B_x^2 \kappa_x  
    -  B_y^2 \kappa_y 
    -  B_z^2 \kappa_z)\mathbf{\nabla}\cdot\mathbf{v}.
    \label{eq:B2}
\end{align} 
To study the conditions under which the magnetic field strength will scale with a power of the local density, we \textit{assume} that the magnetic field strength can be written as
\begin{align}
    B = B_0 \left( \frac{\rho}{\rho_0} \right)^{\kappa},
    \label{eq:condition_0}
\end{align}
where $\kappa$ is constant and $B_0$, $\rho_0$ are the values of $B$ and $\rho$ at $t=0$. By solving the above for $\rho$, substituting that expression into the continuity equation (Eq.~\ref{eq:continuity}), and then multiplying by $2B$, Eq.~\ref{eq:continuity} becomes:
\begin{align}
    \lder[ B^2]{t} & =
    - 2B^2 \kappa \mathbf{\nabla}\cdot\mathbf{v}.
    \label{eq:B_rho}
\end{align}
Eq.~\ref{eq:B2} and Eq.~\ref{eq:B_rho} are consistent only if:
\begin{align}
    B^2 \kappa = B_x^2 \kappa_x +  B_y^2 \kappa_y + B_z^2 \kappa_z.
    \label{eq:kappa}
\end{align}
We will now identify the possible solutions of the above equation.

\subsubsection{Case 1: $B_{x} \gg B_{y}, B_{z}$}
\label{app:1_direction}

In this case, we assume that the magnetic field has one dominant direction, say along the $x$-axis. Then, the magnetic field can be described locally only by the $B_x$ component of the full magnetic field vector.
Hence, Eq.~\ref{eq:kappa} suggests that the magnetic field strength will indeed scale with the local density and that:
\begin{align}
    \kappa = \kappa_x. \label{eq:kappa_1d}
\end{align}
Since we have assumed $\kappa$ to be constant, $\kappa_x$ needs to be constant too and the velocity field is constrained such that (see Eq.~\ref{eq:kappa_x}):
\begin{align}
    \frac{\partial v_x}{\partial x} = \chi, \quad
    \frac{\partial v_y}{\partial y} = \psi, \quad
    \frac{\partial v_z}{\partial z} = \zeta,
\end{align}
where $\chi,\,\psi,\,\zeta$ are constants.
The constant velocity gradients guarantee that the magnetic field will have the same direction as the initial field.

If $B_x$ is the dominant magnetic field component, then the field will scale with $\kappa_x$. If $B_y$ or $B_z$ is the dominant magnetic field component, the field will scale with $\kappa_y$ or $\kappa_z$ respectively.

\subsubsection{Case 2: $B_{x} \sim B_y \gg B_{z}$}
\label{app:2_directions}

Now we assume that the magnetic field has two dominant directions (e.g. along the $x$-axis and $y$-axis), then the magnetic field can be described locally by two components of the full magnetic field vector.
Then, from Eq.~\ref{eq:kappa}, we get that $\kappa=\kappa_x=\kappa_y$, since the magnitude of the magnetic field to be $B^2=B_x^2 + B_y^2$. Adding them together, we get: 
\begin{align}
    \kappa= \frac{1}{2}(\kappa_x + \kappa_y) = 1 - \frac{1}{2}\kappa_z.
    \label{eq:kappa_2d}
\end{align}
In order for $\kappa=\kappa_x=\kappa_y$ and $\kappa$ to be constant the velocity gradients are required to be:
\begin{align}
\frac{\partial v_x}{\partial x} = \frac{\partial v_y}{\partial y} = \chi, \quad
\frac{\partial v_z}{\partial z} = \zeta
\end{align}
where $\chi,\,\zeta$ are constants.

Therefore, if the magnetic field can be described using two components of the full magnetic field vector, its strength will scale with the local density only when the field expands at the same rate along these two directions. If the important components are $B_x$ and $B_y$, then the field will scale as $\kappa=1 - \frac{1}{2}\kappa_z$. For $B_x$ and $B_z$ ($B_y$ and $B_z$) the field will scale with  $\kappa=1 - \frac{1}{2}\kappa_y$ ($\kappa=1 - \frac{1}{2}\kappa_x$).

In general, such a ``2D'' field will not scale with the local density. 
The velocity field restriction is such that it forces the magnetic field to maintain the direction of the total magnetic field vector. As a result, the restriction forces the ``2D'' magnetic field to behave as ``1D'' in the field aligned coordinate system. Thus, in the field aligned system, the magnetic field behaves according to Case 1.

\subsubsection{Case 3: $B_{x} \sim B_y \sim B_{z}$}
\label{app:3_directions}

Now we assume that all the magnetic field components are needed to describe the magnetic field.
Then, from Eq.~\ref{eq:kappa} we get that $\kappa=\kappa_x=\kappa_y=\kappa_z$, in order for the magnitude of the magnetic field to be $B^2=B_x^2 + B_y^2 + B_z^2$. Adding these terms together gives that
\begin{align}
    \kappa = \frac{2}{3}.
    \label{eq:kappa_3d}
\end{align}
In order for $\kappa=\kappa_x=\kappa_y=\kappa_z$ and $\kappa$ to be constant the velocity gradients are required to be:
\begin{align}
\frac{\partial v_x}{\partial x} = \frac{\partial v_y}{\partial y} =\frac{\partial v_z}{\partial z} = \chi. \nonumber
\end{align}
where $\chi$ is constant. 

Therefore, a general magnetic field can scale with the local density with a constant $\kappa$ only if it expands isotropically. 
In general, a ``3D'' field will not scale with the local density. As in Case 2, the velocity field restriction is such that the magnetic field maintains the direction of the total magnetic field vector. Therefore, this restriction makes the magnetic field in the field aligned coordinate system to behave as ``1D''.

\subsection{Velocity field with shearing terms}
\label{app:with_shear}

We now include the shearing terms of the velocity field. Following the same steps as before, we get from the induction equation that:

\begin{align}
    \lder[B^2]{t}  &= 
      2  \mathbf{\nabla}\cdot\mathbf{v}
  \Biggl[ 
    -  B_x^2 \kappa_x 
    -  B_y^2 \kappa_y 
    -  B_z^2 \kappa_z   
  +  B_x B_y \frac{1}{  \mathbf{\nabla}\cdot\mathbf{v} } \left( \frac{\partial v_x}{\partial y} + \frac{\partial v_y}{\partial x} \right) \nonumber \\
  &+ B_x B_z \frac{1}{  \mathbf{\nabla}\cdot\mathbf{v} } \left( \frac{\partial v_x}{\partial z} + \frac{\partial v_z}{\partial x} \right)
  + B_y B_z \frac{1}{  \mathbf{\nabla}\cdot\mathbf{v} } \left( \frac{\partial v_y}{\partial z} + \frac{\partial v_z}{\partial y} \right) 
  \Biggl].
  \label{eq:B2_eul_shear}
\end{align}
Combining with Eq.~\ref{eq:B_rho}, we get the generalized expression of Eq.~\ref{eq:kappa}:
\begin{align}
   B^2 \kappa  &= 
      B_x^2 \kappa_x 
     + B_y^2 \kappa_y 
     + B_z^2 \kappa_z 
  -  B_x B_y \frac{1}{  \mathbf{\nabla}\cdot\mathbf{v} } \left( \frac{\partial v_x}{\partial y}
  + \frac{\partial v_y}{\partial x} \right) \nonumber \\
  &- B_x B_z \frac{1}{  \mathbf{\nabla}\cdot\mathbf{v} } \left( \frac{\partial v_x}{\partial z} + \frac{\partial v_z}{\partial x} \right)
  - B_y B_z \frac{1}{  \mathbf{\nabla}\cdot\mathbf{v} } \left( \frac{\partial v_y}{\partial z} + \frac{\partial v_z}{\partial y} \right).
\end{align}

The above can be written compactly as:
\begin{align}
    \kappa = \frac{1}{B^2} \kappa_{ij} B_i B_j,
    \label{eq:k_shear}
\end{align}
where $i,j$ are indices corresponding to the $x,y,z$ coordinates, and $\kappa_{ij}$ is
\begin{align}
    \kappa_{ij} = 
 \begin{pmatrix}
  \kappa_x  & -\frac{1}{ \mathbf{\nabla}\cdot\mathbf{v} } E_{xy} & -\frac{1}{ \mathbf{\nabla}\cdot\mathbf{v} } E_{xz} \\
  -\frac{1}{ \mathbf{\nabla}\cdot\mathbf{v} } E_{xy} & \kappa_y & -\frac{1}{ \mathbf{\nabla}\cdot\mathbf{v} } E_{yz} \\
  -\frac{1}{ \mathbf{\nabla}\cdot\mathbf{v} } E_{xz} & -\frac{1}{ \mathbf{\nabla}\cdot\mathbf{v} } E_{yz} & \kappa_z
 \end{pmatrix},
 \label{eq:kij}
\end{align}
where $E_{ij}= \frac{1}{2} \left( \frac{\partial v_i}{\partial x_j} + \frac{\partial v_j}{\partial x_i} \right)$ is the strain rate tensor. As $E_{ij}$, $\kappa_{ij}$ is symmetric. Its diagonal elements can also be expressed in terms of the strain rate tensor, so that $\kappa_{ij} = I -  \frac{1}{  \mathbf{\nabla}\cdot\mathbf{v} }\, E_{ij}$, where $I$ is the identity matrix. Therefore, $\kappa_{ij}$ is, as $E_{ij}$, a metric of the deformation of the velocity field.
Eq.~\ref{eq:kappa_1d}, \ref{eq:kappa_2d}, \ref{eq:kappa_3d} are special cases of Eq.~\ref{eq:k_shear}.

The only assumption made to derive Eq.~\ref{eq:k_shear} was that $\kappa$ is constant (in Eq.~\ref{eq:B_rho}). In order for $\kappa$ to be constant, all terms of $\kappa_{ij}$ and $B_i$ have to be independent of position and time.

\subsection{Non-constant $\kappa$}
\label{app:non_constant_kappa}

We now assume that $\kappa$ is not constant, but a general function of $x$, $y$, $z$ and $t$ (i.e. $\kappa(x,y,z,t)$). Substituting Eq.~\ref{eq:condition_0} in Eq.~\ref{eq:continuity} and multiplying by $2B$ we get:
\begin{align}
    \lder[B^2]{t} - 2 \frac{B^2}{\kappa} \ln{\left( \frac{B}{B_0} \right)} \lder[\kappa]{t} = - 2 \kappa B^2 \mathbf{\nabla}\cdot\mathbf{v}.
\end{align}
This is the generalization of Eq.~\ref{eq:B_rho}. We now write Eq.~\ref{eq:B2_eul_shear} compactly as $\lder[B^2]{t} = - 2  \kappa_{ij}B_iB_j  \mathbf{\nabla}\cdot\mathbf{v}$ and substitute it in the previous equation. We get that:
\begin{align}
    \frac{1}{\kappa} \lder[\kappa]{t} = \frac{ \mathbf{\nabla}\cdot\mathbf{v} }{ \ln{\left( \frac{B}{B_0} \right)} } \left( \kappa - \frac{1}{B^2} \kappa_{ij} B_i B_j  \right).
    \label{eq:kappa_equation}
\end{align}
Notice that in order for $\kappa$ to be constant, the term on the right hand side needs to be zero, which gives the previous result for constant $\kappa$ (Eq.~\ref{eq:k_shear}).
If $\kappa$ is changing slowly, so that $\lder[\kappa]{t}\approx0$, then  $\kappa\approx\frac{1}{B^2} \kappa_{ij} B_i B_j$. 
Therefore, Eq.~\ref{eq:k_shear} can describe both the scaling of the magnetic field strength with the local density when $\kappa$ is constant, but also when $\kappa$ is changing slowly in time and deviating slightly from a power law.

\section{Resolution, resistivity and viscosity effects}
\label{app:appendixB}

\begin{figure*}
\centering
\includegraphics[width=\textwidth]{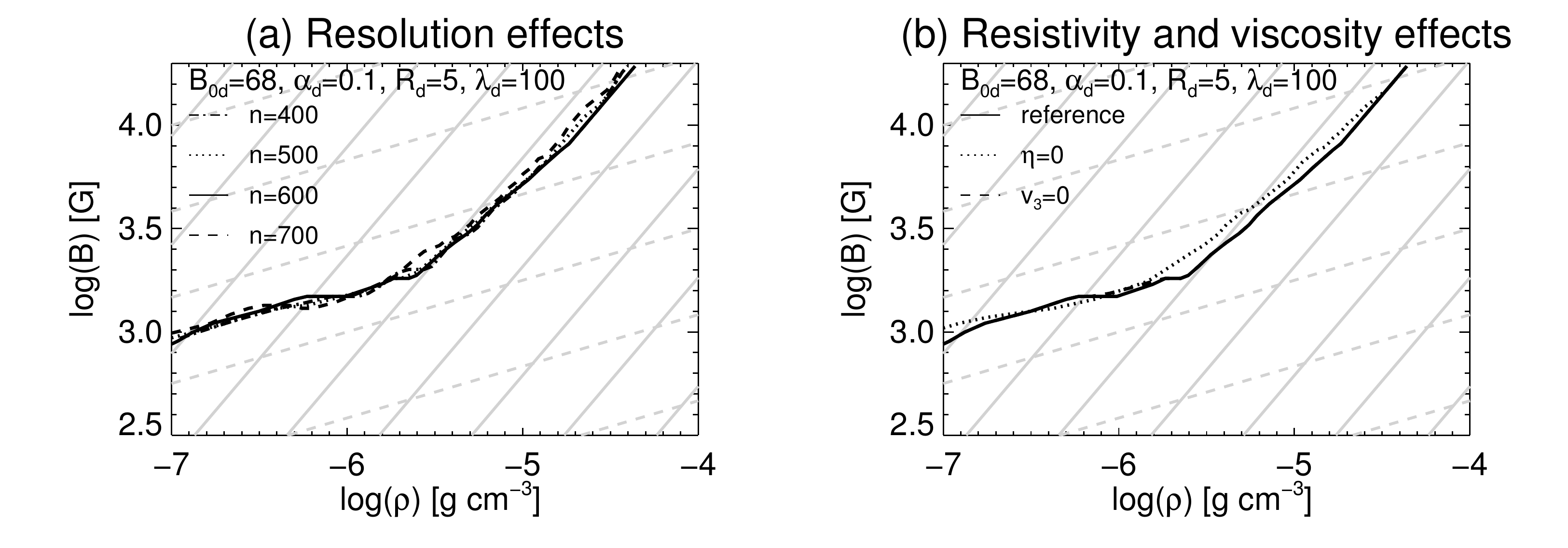}
\caption{
(a) The effect of lower (dotted and dotted-dashed lines) and higher (dashed line) resolution on the scaling curve of case 4 Table~\ref{tab:scaling} (solid line).
(b) The scaling curve of case 4 Table~\ref{tab:scaling} (solid line) in comparison to simulations with the same initial conditions but with $\eta$=0 (dotted line) and $\nu_3=0$ (dashed line).
}
\label{fig:res_adiab}
\end{figure*}

We examine the effect of the resolution on the scaling curves. To do so, we choose case 4 Table~\ref{tab:scaling} to be our reference simulation, as this case is examined in detail in Sec.~\ref{sec:scaling_laws} when explaining the formation the scaling laws. We perform simulations with the same initial conditions and physical domain, and vary the number of grid points. In Fig.~\ref{fig:res_adiab}a, we plot the scaling curve of the reference simulation (solid line, $600^3$ grid points) and compare it with simulations of lower (dotted-dashed line is $400^3$ grid pints and dotted line $500^3$ grid points) and higher (dashed line, $700^3$ grid points) resolution. We find that the scaling curve is minorly affected by the change of the resolution.

We also examine the effect of the viscosity on the scaling curves, by performing a simulation with $\nu_3$=0. We do not set the shock viscosity coefficients ($\nu_1$ and $\nu_2$) to zero to ensure the numerical stability of the simulation. In Fig.~\ref{fig:res_adiab}b, we plot the reference simulation (solid line) and the $\nu_3$=0 simulation (dashed line). The two curved mostly overlap.

Finally, we examine effect of the resistivity on the scaling curves by performing a simulation with no explicit resistivity ($\eta=0$). We plot its scaling curve in Fig.~\ref{fig:res_adiab}b (dotted line). We find that the scaling curve of this simulation is different from the reference simulation by some degree.

The effect of the resistivity on the scaling curves can be estimated analytically by extending the analysis of Appendix~\ref{app:appendix}. Instead of the ideal induction equation, we use the non-ideal induction equation with uniform resistivity:
\begin{align}
    \lder[\mathbf{B}]{t} &= 
    - ( \mathbf{\nabla} \cdot \mathbf{v} ) \textbf{B}
    + ( \textbf{B} \cdot \mathbf{\nabla} ) \textbf{v}
    + \eta \nabla^2 \mathbf{B}. \label{eq:non_ideal_induction}
\end{align}
Using the above in our analysis, Eq.~\ref{eq:B2_eul_shear} becomes:
\begin{align}
    \lder[B^2]{t} = - 2  \kappa_{ij}B_iB_j  \mathbf{\nabla}\cdot\mathbf{v}  
                    + 2 \eta B_i \nabla^2 B_i
\end{align}
Combining the above with Eq.~\ref{eq:B_rho} gives:
\begin{align}
    \kappa = \frac{1}{B^2} \kappa_{ij} B_i B_j 
           - \frac{\eta}{\mathbf{\nabla}\cdot\mathbf{v}} \frac{ B_i \nabla^2 B_i }{ B^2}.
    \label{eq:k_eta}
\end{align}
So, resistivity will have an effect on the value of $\kappa$. The second term of the above equation for the reference simulation is of the order of $0.01-0.1$ during the simulation and this is approximately the difference we find between the solid and the dotted curves in Fig.~\ref{fig:res_adiab}b.

\bibliographystyle{aasjournal}
\bibliography{bibliography}

\clearpage

\end{document}